\documentclass[twocolumn, trackchanges, twocolappendix]{aastex701}

\usepackage{physics}
\usepackage[dvipsnames]{xcolor}

\begin{document}

\shorttitle{Asteroseismology of Low-Mass Pop III Stars} \shortauthors{T. Ferreira et al.}

\title{Red-Giant Asteroseismology of Low-Mass Population III Stars}

\correspondingauthor{Thiago Ferreira dos Santos}
\author[orcid=0000-0003-2059-470X,gname='Thiago',sname='Ferreira dos Santos']{Thiago Ferreira}
\affiliation{Department of Astronomy, Yale University, 219 Prospect St., New Haven, CT 06511, USA}
\email[show]{thiago.dossantos@yale.edu}  

\author[orcid=0000-0003-4456-4863,sname='Bellinger']{Earl P. Bellinger}
\affiliation{Department of Astronomy, Yale University, 219 Prospect St., New Haven, CT 06511, USA}
\email[]{earl.bellinger@yale.edu}

\author[orcid=0000-0002-5794-4286,sname='Farag']{Ebraheem Farag}
\affiliation{Department of Astronomy, Yale University, 219 Prospect St., New Haven, CT 06511, USA}
\email[]{ebraheem.farag@yale.edu}

\author[orcid=0000-0001-8722-1436,sname='Lindsay']{Christopher J. Lindsay}
\affiliation{Department of Astronomy, Yale University, 219 Prospect St., New Haven, CT 06511, USA}
\email[]{christopher.lindsay@yale.edu}

\begin{abstract}

    Low-mass stars from the first epoch of star formation may still persist in the Milky Way and its satellite dwarf galaxies today; however, their detection is confounded by surface pollution from interstellar accretion and internal mixing, which obscure their primordial composition and blur their distinction from second-generation stars. Asteroseismology offers a probe of the internal structure and evolutionary state of stars, and hence may aid in the search for primordial stars. In this second paper of the series, we present the first non-radial adiabatic pulsation analysis of low-mass, metal-free stellar models. We use a $0.85~M_\odot$ red giant as a case study and compare its seismic signatures with those of higher-metallicity models. At the same central hydrogen fractions, Pop III main-sequence models display systematically higher $r_{02} \equiv \delta\nu_{02}/\Delta\nu$ ratio and lower $\Delta\nu$ than metal-enriched analogues, a direct consequence of their larger internal sound speeds and steeper core–envelope stratification. To interpret the structural dependence during giant evolution, we introduce a composite asteroseismic diagnostic, $\psi \equiv \Delta\nu / \Delta\Pi_1$, which traces how metallicity influences the balance between acoustic and buoyancy cavities through its imprint on opacity, core contraction, and mean molecular-weight gradients. Pop III models occupy a distinct locus in the $\psi-\Delta\Pi_1$ plane due to their radiative interiors with lower mean densities and delayed development of core mean molecular weight gradients. We find that asteroseismology is a powerful diagnostic for identifying relic Pop III stars despite potentially polluted surfaces, providing a clear pathway for future searches of the Galaxy's oldest surviving stars with upcoming surveys. 
    
\end{abstract}


\section{Introduction}\label{sec:introduction}

In the absence of metals, radiative cooling is inefficient, leaving primordial gas warm and evolving under opacity-dominated conditions where line cooling saturates, and radiation escapes only diffusively. In this regime, the coupled effects of radiation, turbulence, rotation, and accretion sensitively regulate fragmentation and stellar masses, so that small variations in initial conditions can produce widely different outcomes, limiting our ability to robustly constrain the formation pathways and characteristic masses of Population III stars (Pop III; \citealt{2001ApJ...548...19N, 2016MNRAS.462.1307S, 2017MNRAS.468..418F, 2022ApJ...925...28L}). {These are the first generation of stars formed from primordial Big Bang nucleosynthesis---hydrogen, helium, and trace lithium---and} whilst theoretical advances have provided increasingly detailed predictions for massive Pop III stars ($M \gtrsim 10~M_\odot$), including their radiative feedback \citep{2002A&A...382...28S, 2007ApJ...659..908S, 2024A&A...681A..44S, 2024ApJ...973L..12V}, accretion histories \citep{2009ApJ...701L.133A, 2018MNRAS.481.3278R, 2019MNRAS.483.3592B, 2024ApJ...960L...1N}, and nucleosynthetic yields \citep{2002ApJ...567..532H, 2014ApJ...792...44C, 2017ApJ...848...85J, 2022ApJ...937...61Y, 2024MNRAS.527.5102V, 2024ApJ...967L..22V}, the existence and survivability of low/solar-mass Pop III stars ($M \lesssim 1.2~M_\odot$) is still not understood. Cosmological simulations incorporating detailed thermochemistry and high-resolution hydrodynamics suggest that gravitational fragmentation, driven by disc instabilities and dynamical interactions in primordial mini-haloes, may allow for the formation of even sub-solar mass fragments \citep{1996ApJ...473L..95U, 2015MNRAS.448..568H, 2020ApJ...901...16D, 2020MNRAS.491L..24I, 2021MNRAS.507.1775S, 2022ApJ...927L..12N, 2023MNRAS.524..351K, 2024MNRAS.529.4248T}. If retained and isolated from external enrichment, low-mass Pop III stars could, in principle, survive to the present epoch as pristine stellar relics. {If discovered, these stars would offer unique, local observational windows into the conditions of the early Universe, which would be fundamental for constraining the initial mass function of the first stars \citep{2005ASSL..327..329L, 2016MNRAS.462.1307S}, the onset of cosmic reionisation \citep{2004MNRAS.350...47S, 2006Sci...313..931B, 2010MNRAS.404.1425J, 2025ApJ...994...32V}, the origins of the first heavy elements \citep{2016MNRAS.456.1410S, 2023MNRAS.526.2620V, 2024ApJ...961L..41J}, and the assembly of the first galaxies \citep{2011ARA&A..49..373B, 2013fgu..book.....L, 2020arXiv201002212C}}.

A persistent question, however, remains whether surviving low-mass Pop III stars remain metal-free to the present day. Even if such stars formed with pristine composition, they may acquire metals via accretion from the interstellar medium (ISM) over a few million-year timescales. This possibility is particularly relevant given that their formation likely involves dynamical ejection before significant gas accretion. Within the Bondi-Hoyle-Lyttleton (BHL; \citealt{1939Natur.144.1019H, 1952MNRAS.112..195B}; see also \citealt{2004NewAR..48..843E}) formalism, for instance, the accretion rate onto a star of mass M moving with velocity $v$ through a medium of density $\rho$ and sound speed $c_s$ is given by $\dot{M} = \pi \rho \cdot [(2 G M)^2]/[(v^2 + c_s^2)^{3/2}$. Assuming fiducial parameters representative of the primordial ISM clouds, namely, $M = 0.8~M_\odot$, $v = 20$ km s$^{-1}$, $c_s \approx 10$ km s$^{-1}$, and $n = 1$ cm$^{-3}$ ($\rho \approx 1.67 \times 10^{-24}$ g cm$^{-3}$), we roughly obtain an accretion rate of $\dot{M} \approx 2 \times 10^{11}$ g s$^{-1}$, which integrated over ten billion years, could yield a total accreted mass of $M_{\rm acc} \sim 3 \times 10^{-5}~M_\odot$. If even a modest fraction of this mass consists of metals (e.g., $Z_{\rm ISM} \sim 10^{-4}~Z_\odot$), and assuming that the accreted material is mixed only into the stellar surface convection zone (with typical mass $\sim10^{-3}~M_\odot$), the resulting surface metallicity would be $Z_{\rm surf} \sim 10^{-8} - 10^{-6}$ \citep[see][for in-depth simulations of polluted Pop III stars]{1981A&A....97..280Y, 2015ApJ...808L..47K, 2018PASJ...70...80T}. This range, surprisingly, overlaps with some of the inferred metallicities of the most iron-poor stars observed to date in the Galaxy (e.g., SDSS J102915+172927, with $Z/Z_\odot \approx 10^{-6.1}$; \citealt{2011Natur.477...67C}, SMSS J160540.18-144323.1 with $Z/Z_\odot \approx 10^{-6.2}$; \citealt{2019MNRAS.488L.109N}, and J0023+0307 with $Z/Z_\odot < 10^{-6.6}$; \citealt{2018ApJ...854L..34A}). 

Nevertheless, such a simplified estimate neglects several key processes that can modulate or erase the signatures of external pollution. Stellar winds can expel infalling material or strip away enriched surface layers, particularly in low-mass Pop III stars, where even weak solar-like winds are sufficient to prevent significant metal accretion from the ISM \citep{2017ApJ...844..137T}. Magnetic fields can inhibit accretion by funnelling or deflecting inflows, particularly in magnetically-active low-mass stars, suppressing disc formation or channelling accretion flows (e.g., via magnetic braking or disruption of Keplerian discs; \citealt{2011MNRAS.417.1054S}). Radiative feedback, notably in the pre-main-sequence phase, can heat and disperse surrounding gas, substantially reducing mass accretion rates onto the protostar \citep{2012AIPC.1480...67S}. An additional complication is microscopic diffusion. On the main-sequence, when most long-term accretion is expected, gravitational settling can remove heavy elements from the thin surface convection zones of low-mass stars on timescales compared with their main-sequence lifetimes, potentially erasing any pollution signature. During the giant phases, the deep convective envelope may dilute newly acquired metals through mixing, whilst mass loss and dredge-up further distort the surface composition \citep{2002ApJ...580.1100R, 2002ApJ...571..487V}. Therefore, the uncertain extent of these processes makes the unambiguous identification of low-mass Pop III survivors extremely challenging.

Surface pollution may also arise from internal processes, as discussed in Paper I (\citealt{2026arXiv260216082F}), from convective dredge-up of metals from the stellar interior (particularly following hydrogen-shell merger flashes, {i.e., episodes in which the hydrogen-burning shell penetrates helium-rich layers during late evolutionary phases, triggering convective mixing}). Therefore, the photospheric composition of a primordial star may not faithfully reflect its initial metallicity, and such dual ambiguity between internal dredge-up, external ISM accretion, and/or supernovae enrichment renders surface abundances insufficient to reliably distinguish true Pop III survivors from extremely metal-poor (EMP) second-generation (Pop II) stars. The fundamental challenge, then, lies in probing below the surface, and fortunately, as we show here, asteroseismology offers the means to infer the internal structure of such candidate stars, providing a potentially decisive diagnostic for identifying \emph{bona fide} metal-free survivors. We note that in the (perhaps distant) future, neutrinos could also, in principle, provide insights into the internal processes of Pop III stars; however, their emission timescales may be brief, their nature and flavours may remain poorly constrained, and current detection capabilities may be unlikely to permit unambiguous association with Pop III progenitors, as we discussed in Paper I (see also \citealt{2021MNRAS.508..828N, 2021APh...12802557C}). 

Stellar oscillation spectra are sensitive to microphysics, and particularly the opacity and equation of state \citep{1991ApJ...381..333G, 2022MNRAS.516.5816X, 2023ApJ...942L..38D, 2024ApJ...977....1D, 2024arXiv241001715B}, which differ markedly in metal-free stars. The absence of CNO elements suppresses nuclear burning, thereby slowing energy generation, steepening radiative gradients, and narrowing convective zones. Reduced opacity leads to more compact structures with stiffer cores and steeper Brunt–V{\"a}is{\"a}l{\"a} profiles, reshaping pressure and gravity cavities and altering mixed-mode spectra. These effects are particularly evident in red giant branch stars, where Pop III models show stronger mode coupling, distinct period spacings ($\Delta\Pi_1$), and seismic signatures from acoustic glitches at convective boundaries, {as we demonstrate in this work}. Despite these promising diagnostics, asteroseismic models of Pop III stars remain virtually unexplored. With the precision of {\it Kepler}/K2 \citep{2007CoAst.150..350C, 2015PASP..127.1038C}, the Transiting Exoplanet Survey Satellite (TESS; \citealt{2015JATIS...1a4003R, 2022MNRAS.512.1677S}), and the forthcoming PLAnetary Transits and Oscillations of stars (PLATO; \citealt{2025ExA....59...26R}), seismic studies of low-mass primordial stars may finally become viable.

Although low-mass Population~III stars may persist today in a variety of evolutionary stages ranging from still being on the main sequence to being an advanced white dwarf, our analysis focuses on red-giant models with initial mass $0.85~M_\odot$. Whilst main-sequence stars are far more numerous, their oscillation amplitudes are intrinsically weak, rendering precise asteroseismic detections exceedingly difficult. To date, only an order of a hundred solar-type oscillators have been characterised with high fidelity \citep[e.g.,][]{2011ApJ...743..143H}. This number is expected to rise to $\sim5000$ in the coming years with the advent of the ESA PLATO mission \citep{2025ExA....59...26R}. Empirically, the amplitudes of solar-like oscillations in red giants are known to depend on metallicity, with metal-rich stars exhibiting systematically higher mode amplitudes and longer mode lifetimes, a trend attributed to metallicity-dependent convective-driven and granulation properties (e.g., \citealt{2017A&A...605A...3C, 2018ApJS..236...42Y, 2018A&A...616A..94V}). Nonetheless, given that the anticipated frequency of surviving Pop III main-sequence stars is no greater than one per million objects, their identification through asteroseismology remains exceedingly improbable \citep{2018MNRAS.473.5308M, 2022ApJ...937L...6R}. In contrast, red giants exhibit much larger oscillation amplitudes and are already represented by at least tens of thousands of high-quality seismic targets, which is expected to soon expand into the millions with forthcoming space-based surveys. {Similarly, semi-analytic and cosmological models suggest that surviving low-mass Pop III stars should be exceedingly rare in the Milky Way halo, but potentially more prevalent in low-mass satellite galaxies, with a non-negligible fraction expected to be observed in the red-giant branch phase (e.g, Figure 10 in \citealt{2018MNRAS.473.5308M}; see also \citealt{2016ApJ...820...59K, 2016ApJ...826....9I}). Although the absolute numbers remain highly uncertain, these studies indicate that red-giant Pop III survivors are more observationally accessible than their main-sequence counterparts, further motivating the focus on evolved low-mass models in this work}. {A star of initial mass $0.85~M_\odot$ therefore constitutes an ideal case study, as discussed in Paper I: it is low enough in mass to ensure survival to the present epoch, yet sufficiently evolved to develop mixed-mode spectra that reveal the structure and coupling of the core and envelope with remarkable diagnostic power, whilst also being representative of one of the longest-lived low-mass Pop III stars that could still exist in the Milky Way today.}

In this second paper of our series, we present detailed asteroseismic analyses of low-mass Pop III stars, focusing on a $0.85~M_\odot$ model as a case study. Non-radial adiabatic oscillations ($\ell = 0, 1, 2, 3$) were computed using the {\sc GYRE} {\tt v7.1} oscillation code \citep{2013MNRAS.435.3406T, 2018MNRAS.475..879T}, with evolutionary models generated using Modules for Experiments in Stellar Astrophysics ({\sc MESA} {\tt r24.08.1}) that were presented in the first paper of this series\footnote{MESA {\tt inlists} can be found at \\ \url{https://github.com/thiagofst/PopIII}}. {A brief overview of asteroseismic principles is provided in \S\ref{sec:overview}. In \S\ref{sec:seismic_signatures} and \S\ref{sec:psi}, we examine the global structure of the oscillation spectra of main-sequence and evolved zero-metallicity stars, and explore how their seismic signatures are affected by binarity (\ref{sec:binarity}), interstellar-medium accretion (\ref{sec:accretion}), envelope stripping in higher-metallicity progenitor models (\ref{sec:stripping}), and boundary convective mixing (\ref{sec:convective}); the latter motivated by the results of Paper I.} Particular attention is given to the RGB phases, where solar-like oscillations and mixed modes provide direct constraints on core compactness, convective boundaries, and envelope stratification. {In addition to mapping the oscillation spectra of Pop III stars, we identify seismic signatures that distinguish them from metal-poor Pop II models---subsequent stellar populations formed from gas enriched by the first supernov\ae---even when surface observables are similar, and we introduce a new observable, $\psi$, to quantify this distinction.} Discussions and the conclusions of our results are presented in \S\ref{sec:conclusions}.

\section{Non-Radial Stellar Oscillations}\label{sec:overview}

In non-rotating, non-magnetic stars, spherical symmetry is preserved and oscillation modes of stars can be expanded in spherical harmonics following $Y_\ell^m(\theta,\phi)e^{i\sigma t}$, where $\ell$ denotes the angular degree and ${\sigma = \frak{Re}(\sigma) + i\cdot\frak{Im}(\sigma)}$ is the complex eigenfrequency. {Breaking this symmetry, for instance through rotation or strong magnetic fields, lifts the $m-$degeneracy and causes frequency splitting, with modes of the same $\ell$ but different $m$ acquiring slightly different eigenfrequencies, see, e.g., \citep{1951ApJ...114..373L, 1990MNRAS.242...25G, 2024MNRAS.534.1060H}.} Within such spherical symmetry, frequencies are independent of the azimuthal order $m$. For purely adiabatic radial oscillations, the radial displacement is governed by a second-order Sturm-Liouville type equation%\footnote{{In standard form, this reads} \begin{equation*} \frac{d}{dr}\Big(\Gamma_1 P r^2 \frac{d \xi_r}{dr}\Big) + \left[\sigma^2 \rho r^2 - \frac{d}{dr}(3 \Gamma_1 P)\right] \xi_r = 0, \end{equation*} where $\xi_r(r)$ is the radial displacement, $\Gamma_1$ is the adiabatic exponent, $P$ the pressure, $\rho$ the density, and $\sigma$ the eigenfrequency.} 
yielding the eigenfunction $\xi_{r,n\ell}(r)$. However, when considering non-radial, non-adiabatic oscillations, as implied by the presence of a non-zero imaginary component in $\sigma$, the problem becomes intrinsically more complex, requiring the solution of a coupled sixth-order system of differential equations: {linearised equations for the radial and horizontal displacements ($\xi_r$, $\xi_h$), pressure and density perturbations ($\delta P$, $\delta \rho$), gravitational potential perturbation ($\delta \Phi$), and energy/temperature perturbation ($\delta T$).} For in-depth discussions we refer to \cite{1989nos..book.....U, 2018adaf.book.....B, 2017A&ARv..25....1H, 2018PhDT.......125B}. 

% Mode classification
These oscillations are broadly categorised as pressure (p) or gravity (g) modes, depending on the dominant restoring force. {For p-modes, which are dominant in the outer layers of stars exhibiting oscillations excited by near-surface turbulent convection (so-called solar-like oscillators)}, displacement causes compression, and pressure pushes the material back to its equilibrium point. On the other side, when a parcel is displaced vertically, buoyancy (due to density gradients) tries to restore it. These last are low-frequency g-modes, and are {more} sensitive to the core properties of a star. {Those exhibiting such oscillations, encompassing both solar-type main-sequence stars and red giants, are classified as solar-like oscillators, with the latter displaying substantially larger oscillation amplitudes due to their extended radii and higher luminosities. These oscillations are stochastically excited by convective motions in the outer envelope, offering detailed and insightful diagnostics of the internal structure and processes within the star \citep{2010ApJ...723.1607H, 2013ARA&A..51..353C}.}

{The stochastic excitation by near-surface convection not only determines which modes are present, but also sets the characteristic frequency at which oscillations attain maximal amplitude. This frequency, denoted as $\nu_{\rm max}$, corresponds to modes most efficiently excited and trapped within the stellar envelope, and is closely related to the acoustic cut-off frequency of the atmosphere \citep{1995A&A...293...87K}. Physically, modes with $\nu > \nu_{\rm max}$ are poorly trapped and damped rapidly, whereas those near $\nu_{\rm max}$ attain the largest observable amplitudes. Empirically, $\nu_{\rm max}$ scales with surface gravity and effective temperature as}

\begin{equation}
    \nu_{\rm max} \propto \frac{g}{\sqrt{T_{\rm eff}}} \propto \frac{M / R^2}{\sqrt{T_{\rm eff}}}, \label{eq:numax}
\end{equation}
{where $g$ is the surface gravity, $T_{\rm eff}$ the effective temperature, and $M$ and $R$ the stellar mass and radius. This scaling makes $\nu_{\rm max}$ a robust indicator of global stellar properties and defines the frequency range where solar-like oscillations are most readily observed, where, in combination with the large frequency separation $\Delta \nu$, $\nu_{\rm max}$ forms the basis of widely used asteroseismic scaling relations that allow for precise inference of stellar mass, radius, and evolutionary stage \citep{2011ApJ...743..143H}. Main-sequence stars typically exhibit higher $\nu_{\rm max}$ owing to their stronger surface gravity, whilst red giants show lower $\nu_{\rm max}$, reflecting their expanded envelopes and reduced surface gravities.}

%Characteristic Frequencies and Propagation Zones
The behaviour of oscillation modes is governed by two characteristic frequencies: the Lamb frequency, 

\begin{equation}
    S_\ell^2 = \frac{\ell(\ell+1)c_s^2}{r^2},\label{eq:lamb}
\end{equation}

and the Brunt-V{\"a}is{\"a}l{\"a} (buoyancy) frequency,

\begin{eqnarray}
    N^2 &=& g\left( \frac{1}{\Gamma_1P} \frac{dP}{dr} - \frac{1}{\rho} \frac{d\rho}{dr} \right) \notag \\ 
    &=& \frac{g}{H_P} \left[ \nabla_{\rm ad} - \nabla + \left(\frac{\varphi}{\delta}\right) \nabla_\mu \right],\label{eq:bruntvaisala}
\end{eqnarray}
where $c_s$ is the adiabatic sound speed, $\Gamma_1$ the adiabatic exponent, $g$ the local gravitational acceleration, and $H_P \equiv -\dd r / \dd \ln P$ is the pressure scale height. Temperature gradients are given by $\nabla \equiv \dd \ln T / \dd \ln P$ (the actual gradient) and $\nabla_{\rm ad}$ (the adiabatic gradient). The term $\nabla_\mu \equiv \dd \ln \mu / \dd \ln P$ denotes the mean molecular weight gradient, and the thermodynamic derivatives $\varphi \equiv \left( \partial \ln \rho / \partial \ln \mu \right)_{P, T}$ and $\delta \equiv -\left( \partial \ln \rho / \partial \ln T \right)_{P,\mu}$ account for compositional and thermal buoyancy effects, respectively \citep{2004sipp.book.....H}.  

{These frequencies basically define where modes can propagate within the star}, with zones for each mode type defined by the relations between them and the oscillation frequency $\nu$, i.e., pressure modes propagate where $\nu^2 > \max{(N^2, S_\ell^2)}$, whilst gravity modes are confined to regions where $\nu^2 < \min{(N^2, S_\ell^2)}$. Since $N^2$ is negative in convective regions, g-modes cannot propagate there and become evanescent, i.e., when $N^2 < 0$, the wave equation gives imaginary radial wave-numbers, so that the amplitude of the g-mode decays exponentially to the point of non-propagation. {In this case, g-modes propagate where the radial wave-number $k_r$ is real, and in the asymptotic limit of high-order modes, the local dispersion relation gives \begin{equation*} k_r^2 \simeq \frac{\ell(\ell+1)}{r^2}\left(\frac{N^2}{\sigma^2}-1\right), \end{equation*} where $\sigma$ is the mode frequency and $N^2$ the Brunt–V{\"a}is{\"a}l{\"a} frequency. In convective regions, $N^2 < 0$, so $k_r^2 < 0$ and $k_r$ is imaginary. The radial displacement then behaves as $\xi_r \propto e^{\pm i k_r r} = e^{\mp |k_r| r}$, showing an exponential decay, i.e, g-modes are evanescent in convective zones and do not propagate.}.

{We note that, because solar-like oscillators are stochastically damped by near-surface convection, each mode has a finite lifetime $\tau$, which manifests observationally as a finite line-width $\Gamma$ in the power spectrum $\Gamma \sim (\pi\tau)^{-1}$. Short-lived modes produce broader spectral peaks, whilst long-lived modes appear sharper, with the potential to affect both detectability and measured amplitudes (e.g., \citealt{1989ApJ...341L.103C, 1992ApJ...387..712L, 2009A&A...506...57D}).}

{In evolved stars such as red giants, the contraction of the core, however, leads to an increase in central density, producing pronounced peaks in the Brunt-V{\"a}is{\"a}l{\"a} frequency, whilst the concurrent expansion of the envelope causes the Lamb frequency to decrease. The resulting thin evanescent region between both the g-mode cavity in the core and the p-mode cavity in the envelope allows coupling between these otherwise distinct oscillations, which gives rise to mixed modes \citep{1975PASJ...27..237O, 2017A&ARv..25....1H, 2020ApJ...898..127O, 2020A&A...634A..68P, 2023ApJ...954..152K}. These act as gravity waves in the core and as pressure waves in the envelope, with partial energy tunnelling through the evanescent region, and they provide key diagnostics, in particular, of core structure and composition gradients. The strength of such coupling is quantified by the dimensionless parameter $q$, which depends on the width and stratification of the evanescent zone \citep{1979nos..book.....U, 1992ApJS...80..369B, 2012A&A...540A.143M, 2017A&A...600A...1M}, and the location of the coupling region, approximately where $N \approx S_\ell$, determines how strongly energy can tunnel between the g-mode and p-mode cavities, and particularly governs mixed-mode visibilities, i.e., the amplitude with which these modes appear in the observable oscillation spectrum.}

% Asymptotic Relations and Seismic Diagnostics
Following the asymptotic limit of high radial order ($n \gg \ell$; \citealt{1980ApJS...43..469T}), p-modes exhibit a nearly regular spacing in frequency,

\begin{equation}
    \nu_{n\ell} \simeq \Delta\nu \left(n + \frac{\ell}{2} + \epsilon\right),
\end{equation}
where $\epsilon$ is a phase shift, and the large frequency separation $\Delta\nu$ is described by 

\begin{equation}
    \Delta \nu \approx \left(2\int_{0}^{R}\frac{\dd{r^\prime}}{c_s}\right)^{-1} \propto \left(\frac{M}{R^3}\right)^{1/2} \approx \sqrt{\langle \rho \rangle}, 
    \label{eq:deltanu}
\end{equation}
which reflects a dependency on the mean stellar density (see also \citealt{1967AZh....44..786V, 1986ApJ...306L..37U}). Mode frequencies form patterns in the asymptotic regime, where the eigenfunctions are highly oscillatory and Jeffreys-Wentzel-Kramers-Brillouin (JWKB; \citealt{1924PLMSs23.428J, 1926ZPhy...38..518W, 1926ZPhy...39..828K, 1926CRAS183.24.0000B}) approximation becomes applicable.  

Another diagnostic is provided by the small frequency separation $\delta\nu_{\ell, \ell+2}$ , 

\begin{eqnarray}
    \delta \nu_{\ell, \ell+2} &\equiv& \nu_{n,\ell} - \nu_{n-1,\ell+2} \notag \\ 
    &\approx& -(4\ell+6) \frac{\Delta\nu}{4\pi^2 \nu_{n,\ell}}\int_{0}^{R} \frac{1}{r^\prime} \frac{\dd{c_s}}{\dd{r^\prime}} \dd{r^\prime},
    \label{eq:smallfreqsep}
\end{eqnarray}
which is particularly sensitive to the structure of the stellar core and to chemical composition gradients (e.g., \citep{1995A&A...293...87K, 2003PASA...20..203B}). {Note that, physically, the integral in Equation \ref{eq:smallfreqsep} is dominated by the stellar core, so $\delta\nu_{\ell,\ell+2}$ directly reflects the sound-speed gradient induced by hydrogen depletion and chemical composition gradients. As the core evolves and $X_c$ decreases, the small separation decreases, providing a sensitive measure of the stellar age and internal structure.}

{A related diagnostic that suppresses surface-layer effects and probes the sound-speed gradient in the core, the frequency separation ratio $r_{02}$ is defined by}

\begin{equation}
    {r_{02}(n) \equiv \frac{\delta \nu_{02}(n)}{\delta\nu_1 (n)} = \frac{\nu_{n, 0} - \nu_{n-1, 2}}{\nu_{n, 1} - \nu_{n-1,1}},}\label{eq:r02}
\end{equation}
{where $\nu_{n, \ell}$ is the eigenfrequency of radial order $n$ and degree $\ell$, $\delta \nu_{02}$ is the small separation between $\ell = 0$ and $\ell = 2$ modes, and $\delta \nu_1$ is the large separation of consecutive $\ell = 1$ modes. Notably, $r_{02}$ decreases monotonically with core hydrogen depletion, providing a direct measure of the central hydrogen fraction $X_c$ and main-sequence age (see e.g, \citealt{1988IAUS..123.....C}). For solar-like stars, $r_{02}$ typically varies by $\mathcal{O}(10^{-2})$ across the main sequence, enabling precise constraints on core structure and evolutionary state \citep{2003A&A...411..215R, 2011ApJ...740L...2S}.} {This diagnostic, therefore, provides a clear link between the asymptotic theory of stellar oscillations and an observationally measurable quantity, allowing stellar evolution models to be directly tested against observed frequencies.} {Furthermore, the ratio $r_{02}$ normalises the small separation by the large separation, effectively reducing the influence of poorly modelled near-surface layers, which makes it a robust measure of the sound-speed gradient in the core and also a direct indicator of the evolutionary stage of the star. Mode frequencies, however, can be sensitive to poorly modelled near-surface layers, which introduce systematic offsets when comparing observed and theoretical frequencies \citep{2008ApJ...683L.175K, 2018ApJ...869....8B}}

High-order g-modes show nearly constant period spacing, 

\begin{equation}
    \Delta \Pi_\ell \approx \frac{2\pi^2}{\sqrt{\ell(\ell+1)}} \left( \int_{r_1}^{r_2} \frac{N}{r} \dd{r} \right)^{-1},
    \label{eq:periodspacing}
\end{equation}
with $r_1$ and $r_2$ defining the boundaries of the g-mode propagation cavity, and any deviation from this regularity can indicate sharp composition gradients or mode trapping within the star \citep{1980ApJS...43..469T, 1993MNRAS.265..588D, 2008MNRAS.386.1487M}. {This diagnostic provides an observational probe of the deep stellar interior, as deviations from uniform g-mode period spacing reveal gradients in composition, density, or stratification in the core, allowing models of the internal structure to be also tested against observed oscillation spectra.} {In Equation \ref{eq:periodspacing}, the integral is dominated by the high-density core, so $\Delta \Pi_\ell$ is a sensitive to core stratification. Sharp chemical composition gradients, such as those left by nuclear burning or convective boundaries, can trap modes, producing deviations from uniform period spacing, and observationally, these deviations provide a detailed window into the internal structure of the deep stellar interior \citep{2016MNRAS.457L..59L, 2016A&A...588A..82B, 2019MNRAS.482.1757L, 2023ApJ...954..152K, 2023ApJ...950..165H}.}

\section{Asteroseismology of Low-Mass Population III Stars}\label{sec:seismic_signatures}

\subsection{The Main Sequence}

\begin{figure*}[t]
    \centering
    \includegraphics[width = 0.49\linewidth]{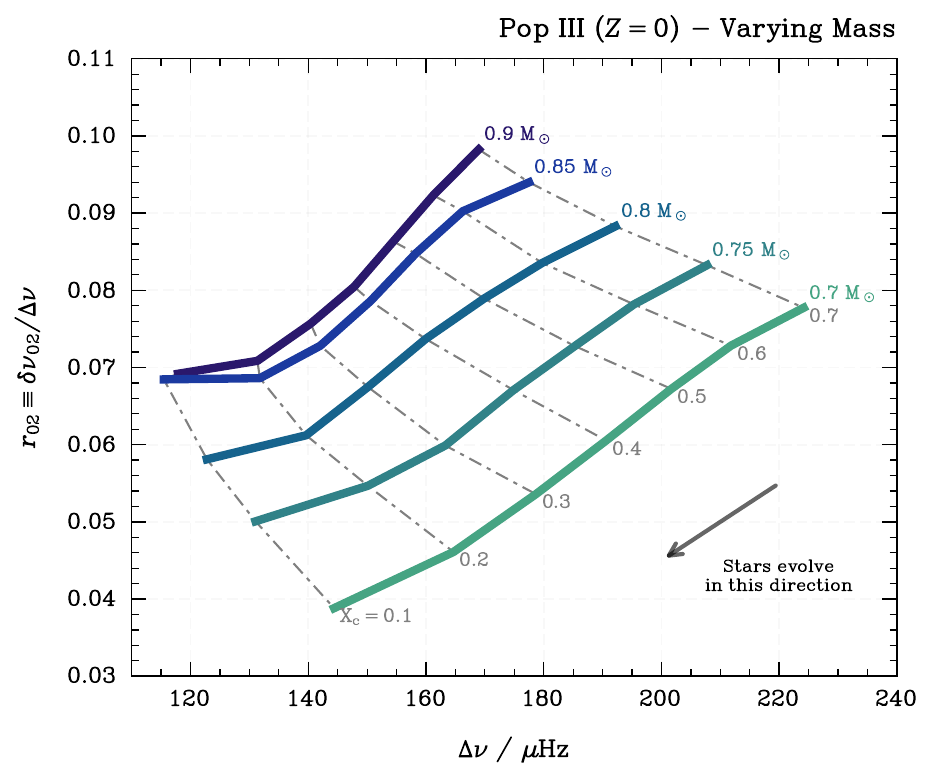} 
    \includegraphics[width = 0.49\linewidth]{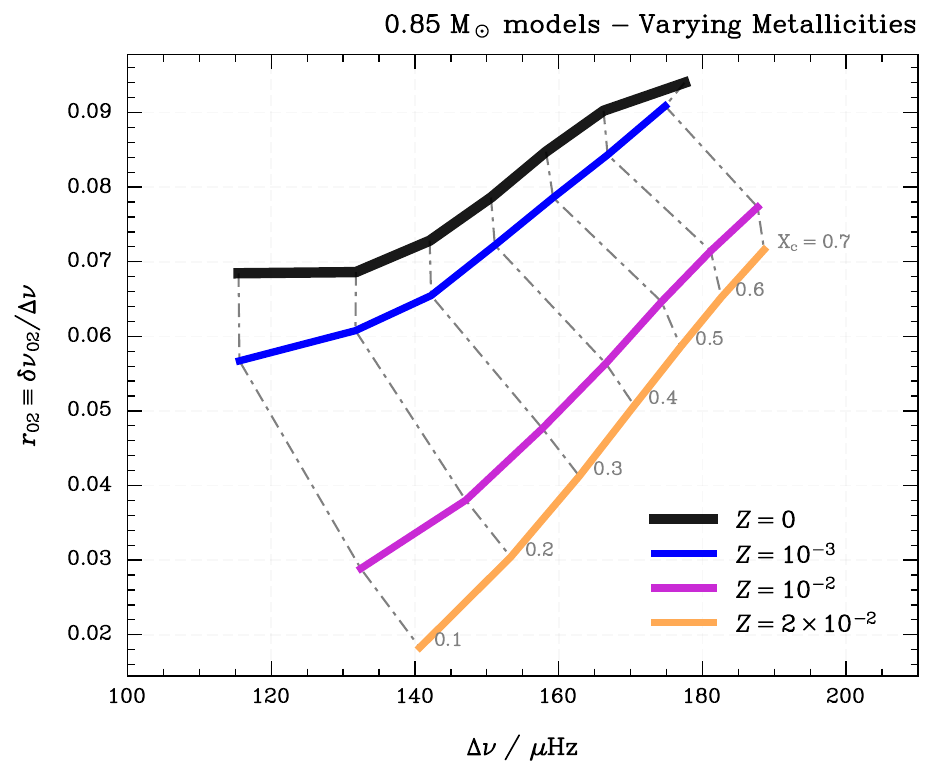}
    \caption{The ratio $r_{02} \equiv \delta\nu_{02}/\Delta\nu$ as a function of the large frequency separation $\Delta\nu$ on the main sequence. The left panel shows results for Pop III stars of varying mass, and the right panel presents models of fixed mass ($0.85~M_\odot$) at different metallicities. Models with metallicities up to $10^{-5}$, including $Z = 0$, are overlaid in the right panel. Grey dash-dotted lines connect evolutionary stages with the same central hydrogen fraction $X_c$, illustrating how asteroseismic properties evolve as the star ages and depletes hydrogen in its core; its mean density decreases due to expansion, resulting in lower values of both $\Delta\nu$ and $r_{02}$. This reflects the slower sound speed and increased sound travel time across the star ({see Equations \ref{eq:deltanu}$-$\ref{eq:r02})}.}
    \label{fig:JCD}
\end{figure*}

Before dredge-up, the absence of metals in Pop III stars lowers their opacity, since metals are the main contributors to bound-bound and bound-free absorption that prevent radiative energy transport \citep{2004sipp.book.....H}. With reduced opacity, radiation escapes more efficiently, whilst the core develops a steeper temperature gradient, channelling energy through narrower thermal pathways. This results in a denser core and a less compact envelope, with higher sound speeds. Additionally, the efficient radiative transport suppresses the development of deep convective envelopes, further reinforcing their strongly radiative and less compact nature. This leads to a smaller large frequency separation ($\Delta\nu$) compared to metal-rich stars of the same mass and central hydrogen fraction $X_c$, as shown in Figure \ref{fig:JCD}. At a fixed $\Delta\nu$, metal-free stars also display elevated values of $r_{02} \equiv \delta\nu_{02} / \Delta\nu$, reflecting stronger core-envelope sound-speed gradients and sharper internal stratification for low-mass stars without convective cores and with minimal molecular weight gradients. {This upward shift is evident in the right panel of Figure \ref{fig:JCD}, where tracks progressively move upward as metallicity decreases from $Z = 10^{-2}$ (bottom) to $Z = 0$ (top).} The $r_{02}$ ratio is particularly sensitive to this internal layering, making it a key probe of structural differences between stellar populations. {Moreover, among Pop III stars, decreasing mass at fixed $X_c$ leads to a slightly higher $r_{02}$, primarily driven by an increase in the small frequency separation $\delta\nu_{02}$ whilst $\Delta\nu$ remains nearly constant, which reflects stronger core-envelope sound-speed gradients and sharper internal stratification in low-mass stars without convective cores and minimal molecular weights gradients.} {The left panel of Figure \ref{fig:JCD} demonstrates such mass dependence, with higher-mass Pop III tracks lying systematically above lower-mass tracks at any given $\Delta\nu$ and at a specific stage of evolution}. In contrast, metal-rich stars have higher opacities and mean molecular weights, which increase radiative trapping and produce broader, less compact global interiors.

\subsection{Subgiants and Red Giants}

\begin{figure}
    \centering
    \includegraphics[width = \linewidth]{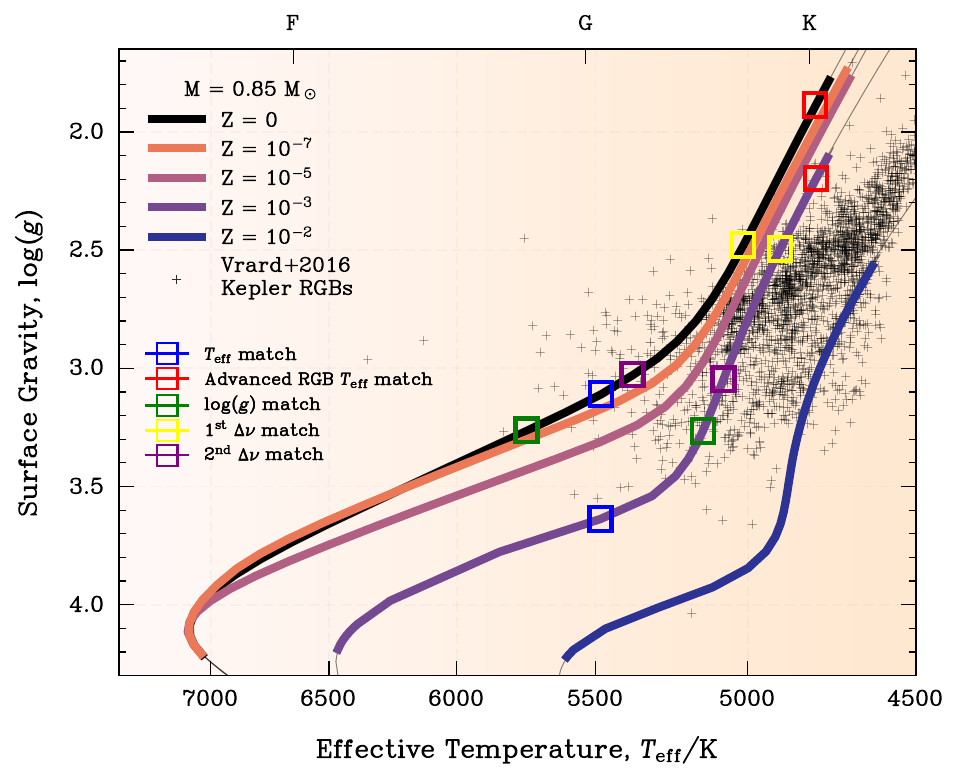}
    \includegraphics[width = \linewidth]{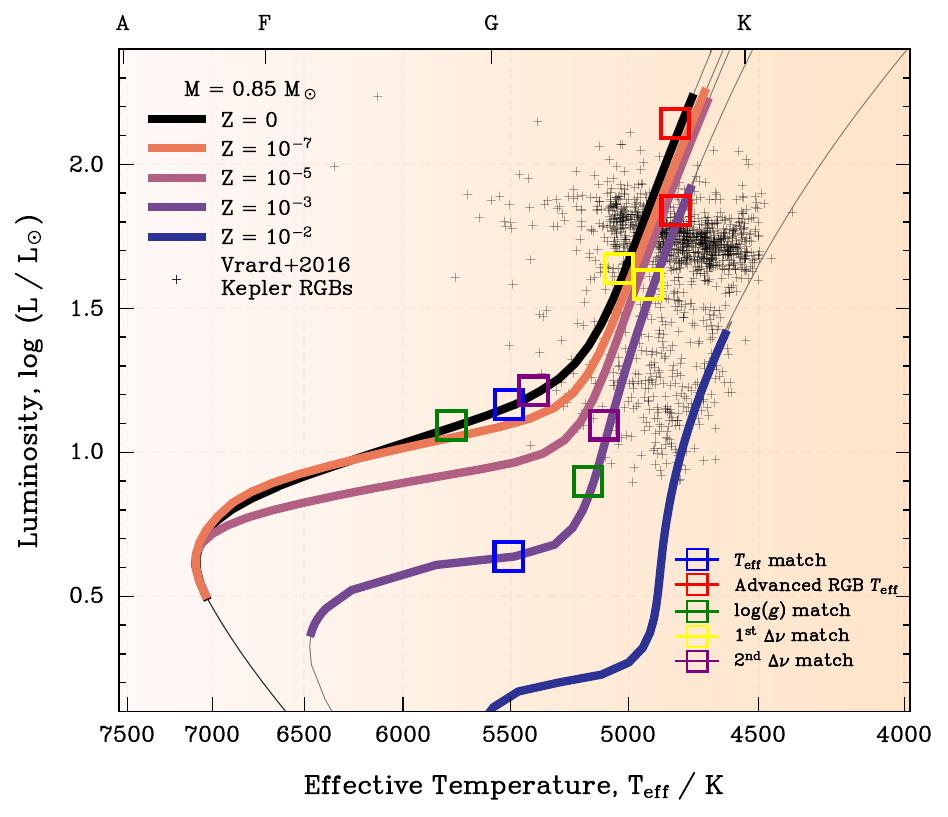}
    \caption{Kiel diagram {(top) and Hertzsprung–Russell diagram (bottom)} for $0.85~M_\odot$ stellar models at distinct metallicities. As metallicity decreases, the evolutionary sequences systematically shift towards higher $T_{\mathrm{eff}}$ at fixed $\log g$, reflecting the reduced envelope opacity and diminished line blanketing in metal-poor atmospheres. Evolution progresses from higher $\log g$ on the main-sequence and subgiant branch towards lower $\log g$ and cooler $T_{\mathrm{eff}}$ along the red giant branch. {Black crosses in the top panels} represent Kepler RGB stars from \cite{2016A&A...588A..87V} for comparison. {The same stars are presented in the bottom panel, however, with luminosities retrieved from the \emph{Gaia} DR3 catalogue \citep{2023A&A...674A...1G}}. Background colours indicate spectral-type domains {for both panels}. The coloured squares correspond to the evolutionary stages of the $Z = 0$ and $Z = 10^{-3}$ models shown in Table \ref{tab:Pop_IIvsIII}. Blue and red squares indicate models matched by effective temperature; green squares indicate those matched by surface gravity; yellow and purple squares correspond to models matched by large frequency separation (see Table \ref{tab:Pop_IIvsIII} and Figure \ref{fig:propagation}).}
    \label{fig:kiel}
\end{figure}

Our discussion now advances to the subgiant and red-giant domains, wherein comparisons between propagation diagrams of Pop II (${Z = 10^{-3}}$) and Pop III models via three matching strategies reveal the significant effect of metallicity on both the internal stellar structure and the associated seismic signatures (see Table \ref{tab:Pop_IIvsIII}). {Figure \ref{fig:kiel} shows the evolutionary tracks in the Kiel diagram, where the three matching criteria are indicated by distinct colours: blue and red for effective temperature-matched pairs, green for surface gravity-matched pairs, and yellow and purple for large frequency separation-matched pairs.} Models were paired by selecting points where a given surface property $\zeta_{\rm PopIII}$ and $\zeta_{\rm PopII}$---specifically, effective temperature, surface gravity, and large frequency separation---differed by no more than a tolerance of $\Delta\zeta = 10^{-3}$. The larger radial extent and higher $N$ in Pop III models enlarge the g-mode cavity and shift the $N \approx S_\ell$ coupling region outward, enhancing mixed-mode visibility and modifying the asymptotic period spacing $\Delta\Pi_1$. Structural differences in core size, density stratification, and evanescent zone width are evident across metallicities and become more pronounced towards the tip of the red giant branch (T-RGB). 

\begin{figure*}[t]
    \centering
    \includegraphics[width = 0.45\linewidth]{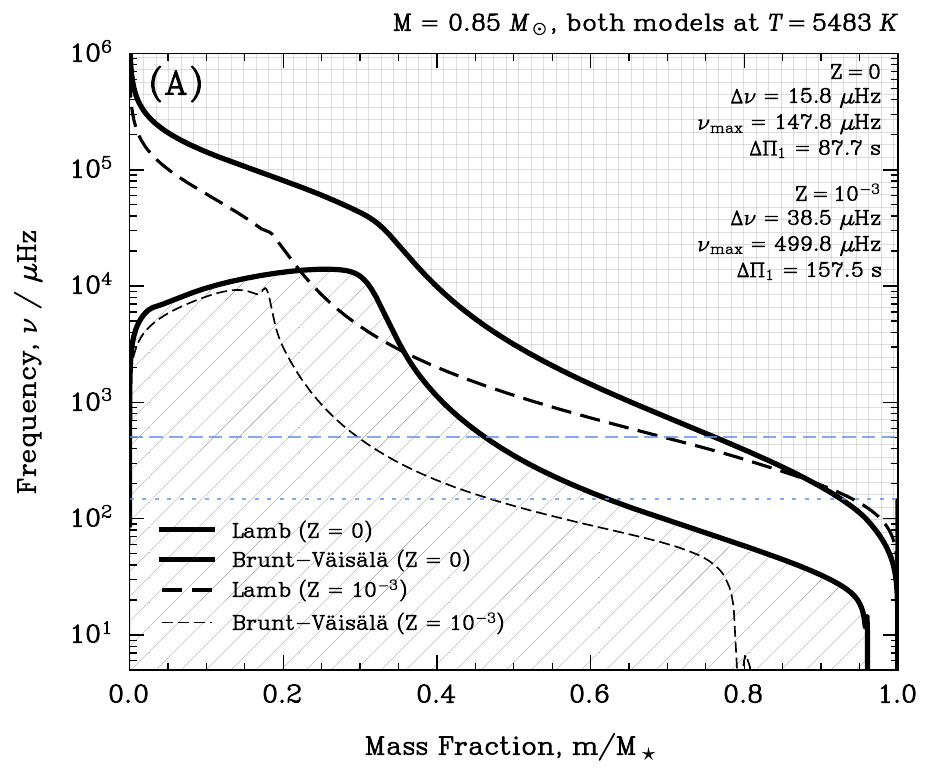}
    \includegraphics[width = 0.45\linewidth]{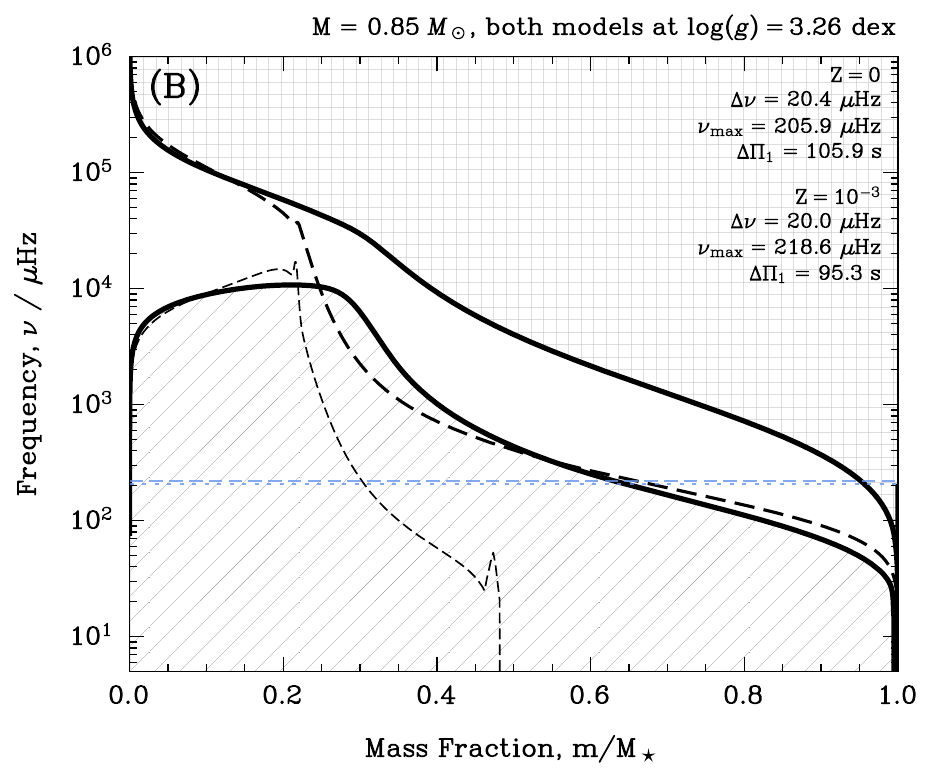}\\
    \includegraphics[width = 0.45\linewidth]{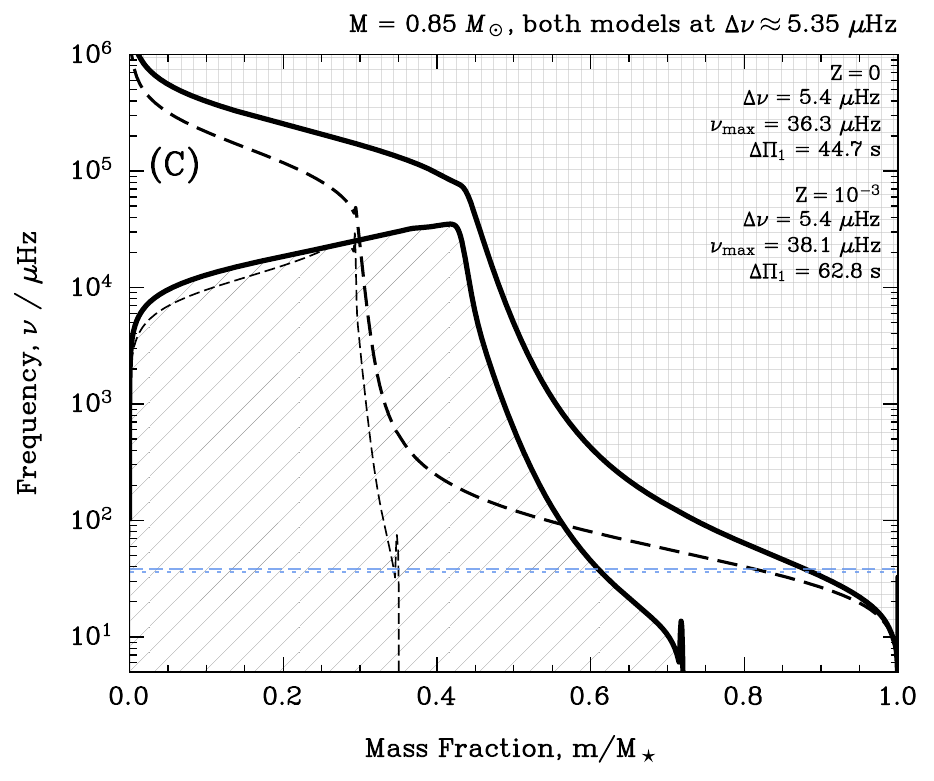}
    \includegraphics[width = 0.45\linewidth]{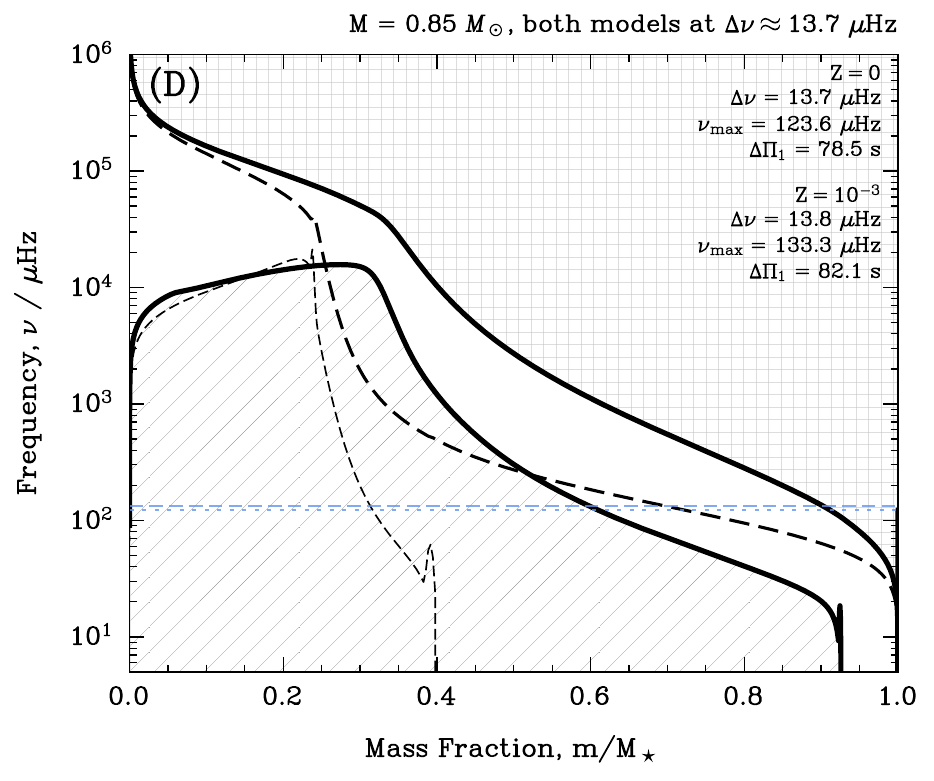}\\
    \includegraphics[width = 0.45\linewidth]{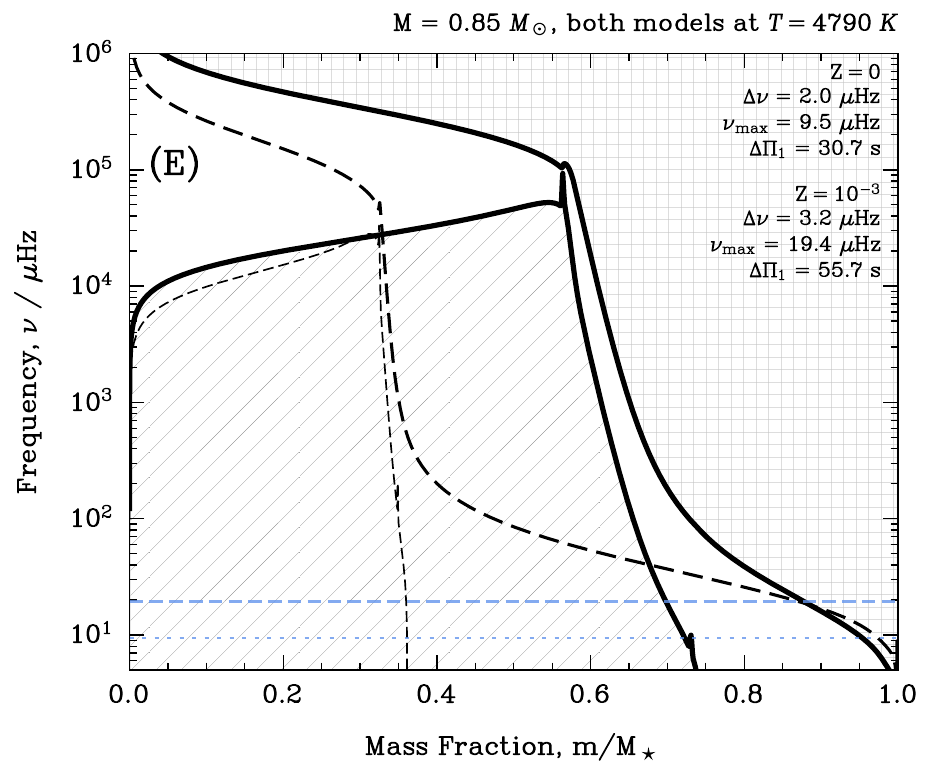}
    \caption{{Propagation diagrams for representative Pop II and Pop III stellar models at three evolutionary stages: subgiant (panel (A); when two models share roughly the same effective temperature), SG-RGB transitions (panels (B), (C), and (D); with both models at roughly the same surface gravity and large frequency separations), and advanced RGB (panel (E); both models at roughly the same effective temperature as well). We refer to Table \ref{tab:Pop_IIvsIII} for a detailed description of the evolutionary phases.} Each panel shows the Brunt-V{\"a}is{\"a}l{\"a} frequency $N$ (solid/dashed for Pop II/Pop III), and the Lamb frequency $S_\ell$ for $\ell = 1$ (solid/dashed) as functions of fractional mass, with $\Delta\nu$, $\nu_{\rm max}$ and $\Delta\Pi_1$ indicated for each case. Dashed/dotted horizontal blue lines indicate $\nu_{\rm max}$ for Pop II/Pop III models, respectively. See also Table \ref{tab:Pop_IIvsIII} summarises the comparisons.}
    \label{fig:propagation}
\end{figure*}

{Figure \ref{fig:propagation} illustrates these structural differences through propagation diagrams at five representative evolutionary stages (panels A--E). Each panel displays the Brunt-V{\"a}is{\"a}l{\"a} frequency $N^{2}$ and Lamb frequencies $S_\ell$ as functions of fractional mass for both Pop II (dashed) and Pop III (solid) models. Panel (A) shows a $T_{\rm eff}-$matched subgiant pair, where the Pop III model exhibits a broader g-mode cavity (region where $N > S_1$) extending to larger radii, despite similar surface temperatures. Panels (B), (D), and (E) correspond to models matched by $\Delta\nu$, where despite similar large frequency separations, the Pop III models show systematically lower $\nu_{\rm max}$ (indicated by horizontal dotted blue lines) and smaller $\Delta\Pi_1$ values, reflecting their extended envelopes and enhanced core stratification. Panel (C) displays a surface gravity-matched pair on the advanced RGB, where the convergence of $N^2$ and $S_1$ at different fractional masses reveals how metallicity reshapes the coupling region between p- and g-mode cavities. Such progressive widening of the g-mode cavity and outward shift of the $N\approx S_1$ crossing point from panels (A) to (E) demonstrates how these structural contrasts amplify as stars ascend the RGB.}

\begin{deluxetable*}{l|c|c|c|c|c}
    \label{tab:Pop_IIvsIII}
    \tablecaption{Comparison of Pop II and Pop III stellar properties at selected evolutionary stages. {\it Notes}: Pop II ($Z = 10^{-3}$) and Pop III ($Z = 0$) stellar models are presented at five representative evolutionary stages: temperature-matched subgiant, surface gravity-matched subgiant, advanced red-giant branch (RGB), and advanced RGB with {two distinct large-frequency separation matches}. The table displays masses, radii, luminosities, helium core masses, effective temperatures, {surface gravities, ages,} and global seismic parameters. Absolute differences equal to, or exceeding 25\%, are highlighted in bold. $^\dagger$Masses differ from the initial $0.85~M_\odot$ owing to mass loss, as detailed in Paper I.}
    \tablehead{
    \colhead{ } & \colhead{$T_{\rm eff}-$matched} & \colhead{$\log(g)-$matched} & \colhead{Advanced RGB} & \colhead{1$^{\rm st}$ $\Delta\nu-$matched} & \colhead{2$^{\rm nd}$ $\Delta\nu-$matched} \\ \hline
    \colhead{} & \colhead{Pop II / Pop III} & \colhead{Pop II / Pop III} & \colhead{Pop II / Pop III} & \colhead{Pop II / Pop III} & \colhead{Pop II / Pop III}} 
    \startdata
    Mass$^\dagger$ ($M_\odot$) & 0.844 / 0.843 & 0.8439 / 0.8439 & 0.837 / 0.818 & 0.840 / 0.836 & 0.8432 / 0.8428\\
    Radius ($R_\odot$) & {\bf 2.31 / 4.25} & 3.55 / 3.56 & {\bf 12.07 / 17.08} & 8.58 / 8.72 & 4.56 / 4.67 \\
    Luminosity ($L_\odot$) & {\bf 4.35 / 14.67} & {\bf 7.93 / 12.40} & {\bf 69.08 / 138.75} & 38.23 / 43.43 & {\bf 12.43 / 16.34} \\
    Helium core mass ($M_\odot$) & {\bf 0.140 / 0.232} & 0.178 / 0.211 & {\bf 0.271 / 0.439} & {\bf 0.24 / 0.33} & 0.20 / 0.24 \\
    Effective temperature (K) & 5483 / 5483 & 5142 / 5744 & 4790 / 4793 & 4899 / 5015 & 5076 / 5374\\
    Surface gravity (dex) & 3.63 / 3.10 & 3.26 / 3.26 & 2.19 / 1.88 & 2.49 / 2.47 & 3.05 / 3.03 \\
    {Age (Gyr)} & {10.78 / 9.78} & {10.997 / 9.723} & {11.23 / 10.03} & {11.20 / 9.94} & {11.08 / 9.80} \\
    \hline 
    $\nu_{\rm max}$ ($\mu$Hz) & {\bf 499.83 / 147.79} & 218.57 / 205.88 & {\bf 19.42 / 9.47} & 38.10 / 36.26 & 133.25 / 123.62 \\
    $\Delta\nu$ ($\mu$Hz) & {\bf 38.50 / 15.77} & 20.04 / 20.40 & {\bf 3.24 / 2.01} & 5.35 / 5.36 & 13.75 / 13.71 \\
    $\Delta\Pi_1$ (s) & {\bf 157.48 / 87.72} & 95.29 / 105.94 & {\bf 55.65 / 30.68} & {\bf 62.82 / 44.71} & 9.69 / 9.02 \\
    \hline
    \hline
    \enddata
\end{deluxetable*}

\subsubsection{Matching by Large Frequency Separation}

For models matched in large frequency separation ({$1^{\rm st}$ match; }$\Delta\nu \approx 5.35~\mu{\rm Hz}$; see Figure \ref{fig:kiel}), corresponding to similar mean densities, Pop II and Pop III models exhibit closely aligned structural properties, yet significant metallicity-driven differences remain apparent in their internal structure and seismic signatures. Despite nearly identical radii ($R_{\rm II}/R_{\rm III} \approx 0.98$) and surface gravities ($\log g_{\rm II}/\log g_{\rm III} \approx 1.007$), the Pop III model is notably hotter ($T_{\rm eff, III} = 5015~{\rm K}$ vs $T_{\rm eff, II} = 4899~{\rm K}$, a $\sim 2.3\%$ difference) and more luminous ($L_{\rm III} \approx 43~L_\odot$ vs $L_{\rm II} \approx 38~L_\odot$, representing a $\sim 12\%$ enhancement), reflecting the reduced opacity and enhanced envelope stratification characteristic of metal-free compositions. The most striking structural difference lies in the helium core mass: the Pop III star exhibits a substantially larger core ($M_{\rm He, III} = 0.33~M_\odot$ vs $M_{\rm He, II} = 0.25~M_\odot$, a $\sim 26\%$ increase), indicative of the prolonged pp-chain burning phase and slower core contraction in the absence of CNO-cycle contributions. 
Seismically, whilst the p-mode cavity sizes remain nearly identical by construction ($\Delta\nu$ differs by only 0.1\%), the g-mode region reveals clear metallicity signatures. The Pop III model shows a markedly smaller asymptotic period spacing ($\Delta\Pi_1 = 45~{\rm s}$ versus $63~{\rm s}$ for the Pop II model, a 40\% reduction), reflecting a smoother Brunt-V{\"a}is{\"a}l{\"a} frequency profile near the core-envelope transition and reduced composition gradient steepness. The frequency of maximum power also differs slightly by $\sim 5\%$ {as a consequence of effective temperature differences (see Equation \ref{eq:numax}).}

These differences highlight that even at fixed $\Delta\nu$, metallicity profoundly shapes the internal stratification, core structure, and coupling between acoustic and gravity-mode cavities. The Pop III stars exhibit more centrally concentrated, larger helium cores with gentler composition gradients, producing distinctly smaller g-mode period spacings that should be detectable in mixed-mode asteroseismic observations of extremely metal-poor subgiants. 

At an earlier evolutionary stage, models matched in large frequency separation at $\Delta\nu \approx 13.7~\mu{\rm Hz}$ also reveal a different manifestation of metallicity effects on subgiant structure. This higher $\Delta\nu$ value corresponds to $R_{\rm II} = 4.6~R_\odot$ vs $R_{\rm III} = 4.7~R_\odot$, characteristic of the early subgiant phase shortly after the exhaustion of core hydrogen. At this stage, the metallicity contrast produces more pronounced photometric differences than observed in the later-stage pair. The Pop III model exhibits a significantly elevated effective temperature ($T_{\rm eff, III} = 5374~{\rm K}$ vs $T_{\rm eff, II} = 5077~{\rm K}$, a $5.5\%$ difference) and substantially higher luminosity ($L_{\rm III} = 16.34~L_\odot$ vs $L_{\rm II} = 12.43~L_\odot$, a $24\%$ enhancement). {These differences in radius, effective temperature, and luminosity are mutually consistent through the well-known Stefan--Boltzmann relation, $L = 4\pi R^2 \sigma T_{\rm eff}^4$, and agree at the few-per-cent level.} {We verified that all other matched model pairs discussed in this section similarly satisfy the fundamental scaling relations within numerical precision.} This amplified luminosity difference, nearly double that of the later-stage pair, reflects the heightened sensitivity of envelope structure to opacity variations during the rapid early subgiant expansion phase.

The helium core mass differential also remains significant, with the Pop III model possessing a $19\%$ larger core ($M_{\rm He, III} = 0.245~M_\odot$ vs $M_{\rm He, II} = 0.198~M_\odot$), smaller than the $35\%$ difference observed at $\Delta\nu \approx 5.35~\mu{\rm Hz}$. This indicates that core mass disparities between Pop II and Pop III stars increase as they ascend the subgiant branch, {and that the differences arise because Pop III stars sustain slightly more efficient hydrogen-shell burning, which deposits helium onto the core at a faster rate. At the same time, the stellar envelope adjusts more slowly to the expanding core, allowing the core to grow larger before structural readjustment occurs.}

Seismically, this early-subgiant pair exhibits remarkably similar properties despite the large photometric differences. The frequency of maximum power differs by $7.8\%$ ($\nu_{\rm max, II} = 133~\mu{\rm Hz}$ vs $\nu_{\rm max, III} = 124~\mu{\rm Hz}$), whilst the asymptotic period spacing shows only modest divergence ($\Delta\Pi_1 = 82~{\rm s}$ vs $79~{\rm s}$, a $4.6\%$ difference). This muted seismic contrast, which is substantially smaller than the $29\%$ $\Delta\Pi_1$ difference at lower $\Delta\nu$ in the RGB pair, indicates that the g-mode cavity structure is less sensitive to metallicity effects at earlier evolutionary phases when composition gradients are still developing, and the hydrogen-burning shell is closer to the core.

Comparing the two $\Delta\nu$-matched pairs also reveals a clear evolutionary progression: at higher mean densities (early subgiant), metallicity primarily affects surface properties and luminosity ($24\%$ difference), whilst internal structure remains relatively similar ($\Delta\Pi_1$ differs by only $4.6\%$). As stars expand and density decreases (late subgiant), photometric differences are moderate (luminosity differs by $14\%$), but internal stratification diverges dramatically ($\Delta\Pi_1$ differs by $29\%$), which suggests that mixed-mode asteroseismology becomes increasingly powerful for metallicity discrimination as low-mass stars ascend the subgiant branch and develop more pronounced core-envelope decoupling.

\subsubsection{Matching by Effective Temperature}

For temperature-matched models ($\Delta{\rm T_{eff}} = 0.01\%$ at $\approx 5480~\mathrm{K}$), the Pop III model has a significantly larger radius, $4.2~R_{\odot}$ compared to $2.3~R_{\odot}$ for Pop II, with radii ratio $R_{\rm II}/R_{\rm III} \approx 0.55$. This enlarged state results from the drastically reduced CNO opacity in the metal-free star, which lowers radiative opacity and, therefore, the internal temperature gradient, allowing hydrostatic equilibrium at lower densities and inflating the envelope. Consequently, the acoustic (p-mode) cavities are systematically extended. 

Moreover, in Pop III models, the helium-core mass is $\approx 40\%$ larger than in Pop II stars ($0.23~M_\odot$ vs. $0.14~M_\odot$), owing to the dominance of the pp chain in the absence of CNO catalysts, which alters core contraction and chemical stratification; these structural differences consequently shape the Brunt-V{\"a}is{\"a}l{\"a} ($N$) frequency profile, with Pop II stars developing steeper composition gradients near the core boundary due to CNO-processed material ($\nabla_\mu \neq 0$), sharply increasing $N$, whereas Pop III models maintain smoother $N$ profiles consistent with pp-chain dominance. This contrast also leads to a pronounced difference in asymptotic g-mode period spacing, with $\Delta \Pi_{1,{\rm II}} / \Delta \Pi_{1,{\rm III}} \approx 1.8$ matching models at similar effective temperatures. Furthermore, global seismic parameters differ markedly: frequency of maximum power $\nu_{\max}$ and large frequency separation $\Delta \nu$ exhibit ratios $\nu_{\max,{\rm II}}/\nu_{\max,{\rm III}} = 3.4$ and $\Delta \nu_{\rm II}/\Delta \nu_{\rm III} = 2.4$, reflecting the impact of metallicity on mean density, surface gravity, and temperature structure. 

During advanced RGB evolution, we also compare models matched by effective temperature ($\approx4790~K$). At this stage, the Pop III star exhibits a considerably larger radius of $17~R_\odot$ compared to $12~R_\odot$ for the Pop II star, accompanied by a factor-of-two higher luminosity (140 vs. 70 $L_\odot$). The helium core mass difference remains substantial, with Pop III exhibiting a 38\% larger core ($0.43~M_\odot$ vs. $0.27~M_\odot$ in Pop II), signifying continued divergent nuclear burning and core growth histories governed by metallicity-dependent processes. Asteroseismic parameters reinforce these structural contrasts: the frequency of maximum power $\nu_{\rm max}$ in Pop II is approximately twice that of Pop III (19~vs. 9.5~$\mu$Hz), and the large frequency separation $\Delta\nu$ is 1.6 times greater (3.2 vs. 2.0~$\mu$Hz). Furthermore, the asymptotic g-mode period spacing $\Delta\Pi_1$ shows an even larger disparity, with Pop II at 56~s and Pop III at 31~s (a ratio near 1.8), indicative of markedly different buoyancy cavity structures shaped by internal composition gradients and core compactness. These seismic diagnostics demonstrate the persistent influence of metallicity well into advanced RGB phases, where Pop III stars maintain more inflated envelopes, larger helium cores, and correspondingly distinct oscillation spectra, suggesting that asteroseismology remains a powerful discriminator of primordial versus metal-enriched stars even at late evolutionary stages, {with Figure \ref{fig:propagation} likely} revealing extended p-mode cavities and altered g-mode trapping consistent with the measured frequency ratios.

\subsubsection{Matching by Surface Gravity}

When the Pop II and Pop III models are compared at matched surface gravity ($\Delta\log(g) = 0.06\%$ at approximately 3.26 dex, they display broadly similar seismic properties yet remain clearly distinct in their global observables. {Their large frequency separations are fairly comparable ($\Delta\nu = 20.04$ and $20.40~\mu$Hz) whilst their $\nu_{\rm max}$ values differ by roughly $\sim6\%$ (219 and 206~$\mu$Hz), as well as their asymptotic g-mode period spacing by $\sim11\%$ (95~s vs. 106~s).} However, their effective temperatures and luminosities differ by roughly 90\% and 36\%, respectively; differences far exceeding any plausible observational degeneracy. {We further note that the observed $\nu_{\rm max}$ may depart non-negligibly from the simple scaling relation prediction, particularly at low-metallicities regimes, as highlighted in recent analyses of metal-poor stellar populations (e.g., \citealt{2025ApJ...989..189L}), with such deviations potentially introducing additional systematic uncertainties when interpreting $\nu_{\rm max}$ differences between Pop II and Pop III models.} Therefore, whilst the internal structures implied by the seismic parameters converge at fixed $\log g$, the surface properties remain so dissimilar that Pop II and Pop III subgiants could be readily distinguished through photometric or spectroscopic means.

\section{$\psi$}\label{sec:psi}

Here we present an asteroseismic diagnostic for identifying Pop III red giants. For seismic analyses, the large frequency separation, $\Delta\nu$, scales with the mean density of the stellar envelope, whilst the dipole gravity-mode period spacing, $\Delta\Pi_1$, is set by the integrated Brunt–V{\"a}is{\"a}l{\"a} frequency in the radiative core, providing complementary probes of envelope and core structure (see \citealt{2011Natur.471..608B, 2012A&A...540A.143M}, and references therein). Their ratio, $\psi \equiv \Delta\nu/\Delta\Pi_1$, links the acoustic and buoyancy cavities, acting as a global structural tracer sensitive to both mean density and core stratification; as shown in the previous section, at a fixed $\Delta\nu$, the value of $\Delta\Pi_1$ can differ significantly, reflecting variations in core structure. As such, $\psi$ may help distinguish stellar populations of different metallicities. On one hand, the large frequency separation, $\Delta\nu$, scales with the square root of mean density and is influenced by the mean molecular weight (cf., Equation \ref{eq:deltanu}). Low-metallicity stars have lower $\mu$ and higher sound speeds, resulting in larger $\Delta\nu$ for similar structures. The period spacing $\Delta\Pi_1$ reflects the structure of the radiative interior and is sensitive to core stratification, particularly gradients in $\mu$ and entropy (cf., Equation \ref{eq:periodspacing}), and as stars evolve off the main-sequence, $\Delta\Pi_1$ decreases due to core contraction and steeper chemical gradients, which amplify the Brunt–V{\"a}is{\"a}l{\"a} frequency.

\begin{figure}[t]
    \centering
    \includegraphics[width = \linewidth]{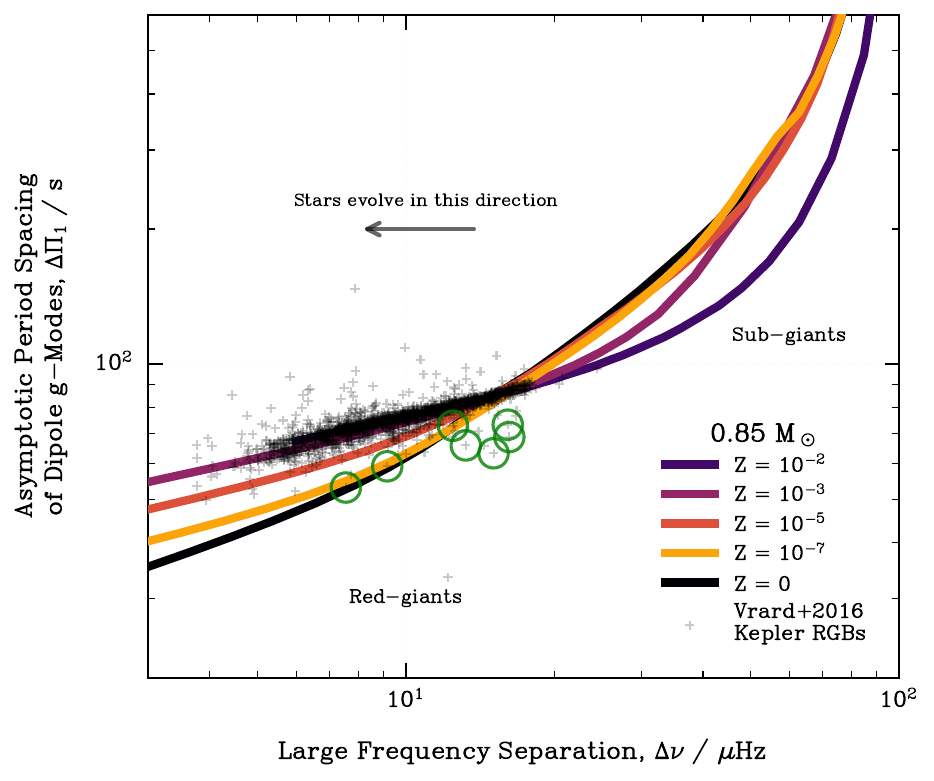}
    \caption{Period spacing $\Delta\Pi_1$ along the subgiant and red giant branch as a function of the frequency spacing $\Delta\nu$ for stellar models of varying metallicity. {KIC stars 10152909, 5696938, 9326896, 9474201, 9945389, 7509923, and 6273090 are highlighted in green as discussed in Section \ref{sec:binarity}.} {We note that in the $\psi-\Delta\Pi_1$ plane (Figure \ref{fig:psi}), the slope of the evolutionary track along the subgiant and red giant branch correspond to approximately the inverse of $\psi$, i.e., $\dd{\Delta\Pi_1}/\dd{\Delta\nu} \sim 1/\psi$ for quasi-linear sections, which provides a visual interpretation of $\psi$ as the gradient of the track, linking steeper slopes to weaker p-g coupling and flatter slopes to stronger coupling.}}
    \label{fig:DeltaNuDeltaPi1}
\end{figure}

At low but non-zero metallicity ($Z \sim 10^{-3}$), stars have lower opacity, leading to compact envelopes and faster core contraction. This produces higher $\Delta\nu$, lower $\Delta\Pi_1$ (see Figure \ref{fig:DeltaNuDeltaPi1}), and thus larger $\psi$. However, at zero metallicity ($Z = 0$), the absence of CNO elements slows core evolution and the development of $\mu$-gradients, resulting in a less peaked Brunt–V{\"a}is{\"a}l{\"a} frequency and comparatively higher $\Delta\Pi_1$. As a result, whilst metal-poor stars show higher $\psi$ than solar-metallicity stars, zero-metallicity models display a {reduction} in $\psi$ near evolutionary turning points, despite high $\Delta\nu$. This is evident in the $\psi-\Delta\Pi_1$ evolutionary tracks, where $\psi$ {exhibits a turnover (knee), after which it declines}, with the maximum value depending on metallicity (see e.g., Figure \ref{fig:psi}).

{The metallicity-dependent morphology of the $\psi-\Delta\Pi_1$ tracks indicates that $\psi$ may serve} as a valuable seismic diagnostic of primordial chemical composition, with the morphology and peak location of the $\psi-\Delta\Pi_1$ relation offering potential signatures of extremely low or even zero metallicity. Given that both $\Delta\nu$ and $\Delta\Pi_1$ are directly accessible from space-based photometry, $\psi$ can be robustly estimated for a broad population of evolved stars. A comparison with \emph{Kepler} RGB stars from \citet{2016A&A...588A..87V} is shown in Figure \ref{fig:psi}. When combined with spectroscopic constraints on the surface [Fe/H], this diagnostic may aid in identifying post-main-sequence candidates for Pop III stellar remnants in future dedicated observational studies.

\begin{figure}
    \centering
    \includegraphics[width = \linewidth]{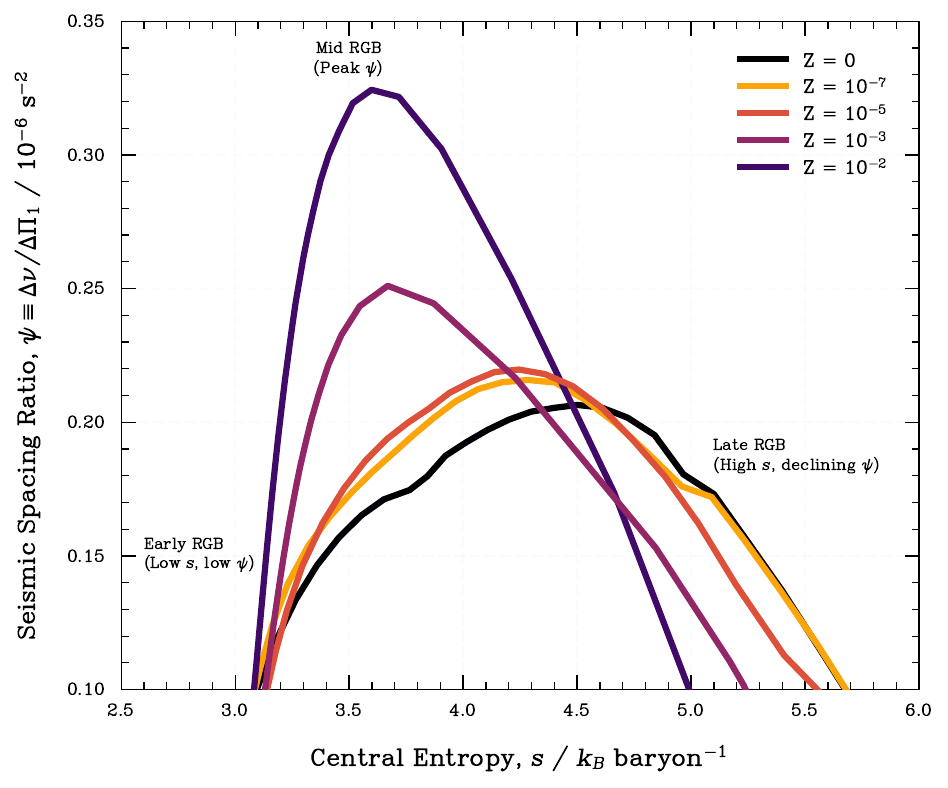}
    \caption{{Seismic spacing ratio $\psi \equiv \Delta\nu / \Delta\Pi_1$ as a function of central entropy $s$ ($k_B$ per baryon) for $0.85~M_\odot$ models across different metallicities. Metal-free models show broader, lower-amplitude peaks due to slower entropy evolution and less efficient CNO burning, whilst metal-enriched models (e.g., $Z = 10^{-2}$) exhibit sharper, higher peaks from steeper temperature gradients and enhanced opacity.}}
    \label{fig:entropy}
\end{figure}

\begin{figure*}[t]
    \centering
    \includegraphics[width = 0.49\linewidth]{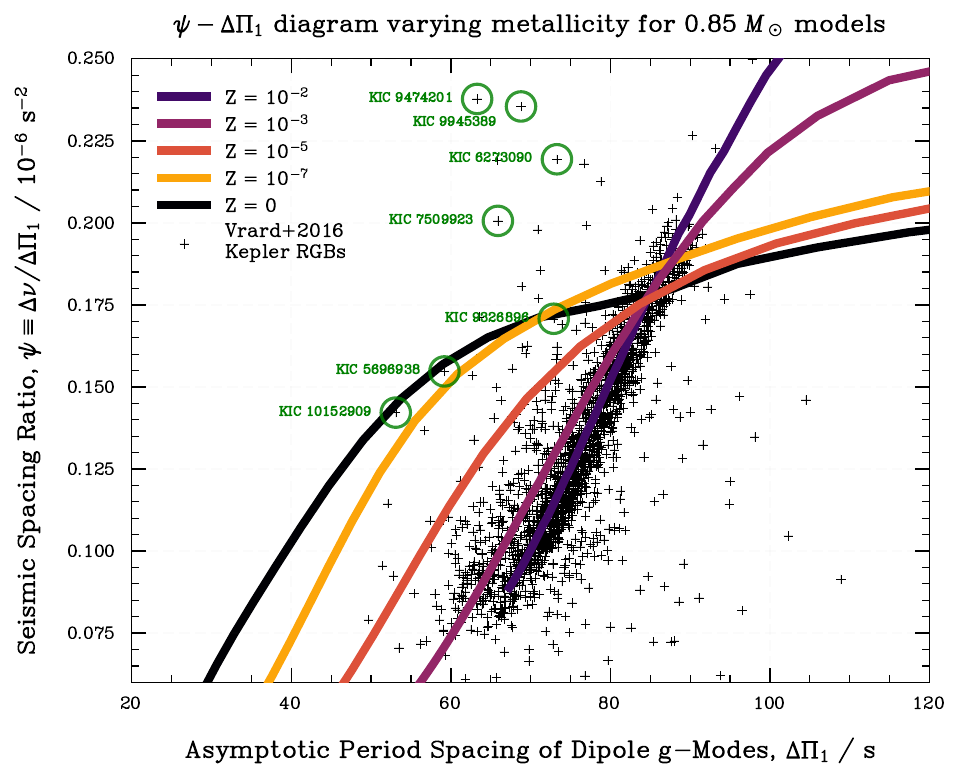}
    \includegraphics[width = 0.49\linewidth]{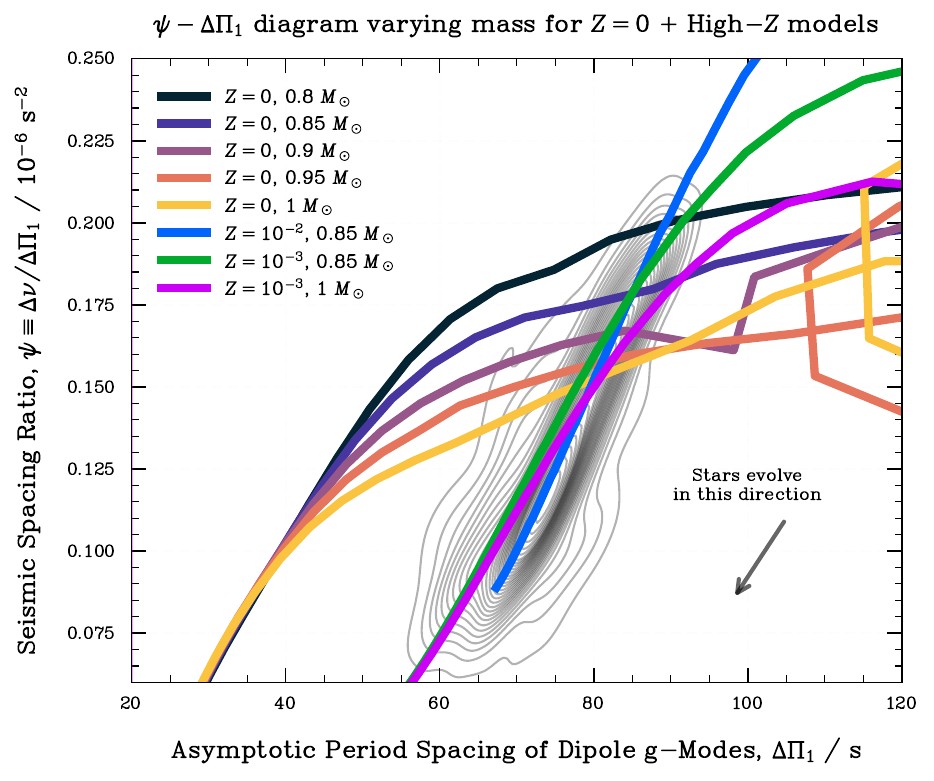}
    \caption{Seismic spacing ratio $\psi \equiv \Delta\nu/\Delta\Pi_1$ as a function of $\Delta\Pi_1$ for $0.85~M_\odot$ models (left). Measurements of RGB stars observed by {\it Kepler} as listed by \cite{2016A&A...588A..87V} are overlaid, and closely follow the solar-metallicity evolutionary track. {KIC stars 10152909, 5696938, 9326896, 9474201, 9945389, 7509923, and 6273090 are highlighted in green as discussed in Section \ref{sec:binarity}.} {\it Right}: As in the left panel, but showing models of varying mass at zero-metallicity (Pop III) and, for comparison, Pop II models; the observed Kepler RGB distribution is indicated by a density contour.}
    \label{fig:psi}
\end{figure*}

The physical origin of these trends can be understood by examining how the seismic spacing ratio $\psi \equiv \Delta\nu / \Delta\Pi_1$ depends on helium core mass and central entropy across models of different metallicity \citep{2013ApJ...765L..41S}. {Central entropy, $s$, is a thermodynamic measure of the core's disorder and thermal content \citep{2004sipp.book.....H}, defined as the entropy per baryon; for a non-degenerate ideal gas, $s \sim k_B (5/2 + \ln(T^{3/2}/\rho))$, where $T$ is temperature and $\rho$ is density, but in degenerate red giant cores, entropy is lower and less sensitive to temperature due to quantum effects (e.g., \citealt{2018A&A...609A..95B} for entropy proxy inversions in asteroseismology; \citealt{2020MNRAS.492.5940H} for the influence of specific entropy in red giant structural evolution). As stars ascend the red giant branch, shell hydrogen burning increases core temperature and entropy, driving a non-monotonic evolution in $\psi$, i.e., at low $s$, the compact, degenerate core yields high $\Delta\nu$ (from elevated mean density) and low $\Delta\Pi_1$ (from strong stratification), resulting in smallest $\psi$; as $s$ rises, core expansion lifts degeneracy, $\Delta\nu$ decreases, $\Delta\Pi_1$ increases, and $\psi$ peaks at intermediate $s$, marking a balance between expansion and buoyancy weakening; at high $s$, the extended core causes $\Delta\nu$ to fall further whilst $\Delta\Pi_1$ rises rapidly, leading to declining $\psi$.} {As stars evolve and their helium cores grow, the central entropy decreases due to core contraction and increasing degeneracy---a process that regulates phenomena such as the mirror principle and the red-giant bump \citep{2020MNRAS.492.5940H}.} In metal-poor and particularly metal-free stars, the slower development of $\mu$-gradients and the more gradual entropy decline result in a broader range of $\Delta\Pi_1$ and a characteristic suppression of $\psi$ at fixed core mass. This behaviour reflects the {slower development of mean molecular weight gradients in the core} and the less efficient CNO burning in Pop III stars, which in turn moderates the growth of the Brunt-V{\"a}is{\"a}l{\"a} frequency and alters the structure of the g-mode cavity. {Metallicity fundamentally modulates this behaviour, with lower opacity in Pop III stars leading to hotter, less stratified cores, producing broader, lower peaks in $\psi(s)$, whilst higher opacity in metal-rich stars creates steeper gradients and sharper, higher peaks, encoding the efficiency of energy transport and degeneracy evolution.} At a given helium core mass, Pop III models occupy a distinct locus in the $\psi-s$ plane compared to their metal-rich counterparts, with lower $\psi$ and higher central entropy {(see Figure \ref{fig:entropy})}. This separation is most pronounced near the onset of core helium burning, where the connection between core mass, entropy, and metallicity imprints a unique seismic signature.

{In the right panel of Figure \ref{fig:psi}, we illustrate the impact of varying stellar mass and metallicity on the $\psi-\Delta\Pi_1$ plane. For $Z = 0$ models, increasing the stellar mass systematically shifts the evolutionary tracks downward in the $\psi-\Delta\Pi_1$ diagram, with higher-mass models exhibiting larger values of $\psi$ at a given period spacing---particularly beyond the subgiant phase---whilst tracks converge during the early RGB phases. In contrast, high-metallicity models (shown in red, blue and green) occupy a distinct locus, characterised by steeper tracks and smaller $\psi$ values at a given $\Delta\Pi_1$, even during early RGB phases. The clear separation between the $Z = 0$ and high-$Z$ tracks, across the mass range considered, demonstrates the robustness of $\psi$ as a diagnostic for distinguishing primordial stars from their metal-enriched counterparts. Background contours, representing the observed distribution of known RGB stars from \cite{2016A&A...588A..87V}, further highlight that the $Z = 0$ models occupy a region of parameter space not populated by typical field stars, reinforcing the potential of asteroseismology for identifying Pop III candidates.}

\subsection{Influence of Binarity }\label{sec:binarity}

Although our models assume single stellar evolution, processes such as mass transfer or mergers can result in an altered core mass and thus introduce additional scatter in $\psi$ (e.g., \citealt{2021MNRAS.508.1618R, 2022A&A...659A.106D}). Interestingly, few stars fall along the $Z = 0$ model lines, and none appear to be the product of single-star evolution, with notable examples including KIC~10152909 ($\Delta\Pi_1 = 53.1~\mathrm{s}$, $\Delta\nu = 7.55~\mu{\rm Hz}$), which may be a rapidly rotating red giant \citep{2020A&A...639A..63G}, KIC~5696938 ($\Delta\Pi_1 = 59.2~\mathrm{s}$, $\Delta\nu = 9.16~\mu{\rm Hz}$) and KIC~9326896 ($\Delta\Pi_1 = 72.9~\mathrm{s}$, $\Delta\nu = 12.45~\mu{\rm Hz}$), which are both likely by-products of mass transfer in binary systems \citep{2022A&A...659A.106D}. These stars, all with masses greater than $1.2~M_\odot$, indicate deviations from standard single-star evolutionary pathways. 

It is worthy of consideration whether a low-mass ($M<1~M_\odot$) binary merger product can also masquerade as a Pop III star; however, forming a present-day $\sim0.85~M_\odot$ giant via coalescence requires two progenitors with initial masses $<0.85~M_\odot$ that both survive and interact only after nearly a Hubble time on the main-sequence. Yet stars with $M \lesssim 0.8~M_\odot$ have main-sequence lifetimes comparable to or exceeding the age of the Universe, and would not naturally undergo envelope expansion, tidal coupling, or unstable mass transfer early enough to drive a merger. Consequently, assembling such a system demands finely tuned orbital evolution or non-standard angular-momentum loss channels, making this scenario physically disfavoured. We explore seismic effects of envelope stripping in Section \ref{sec:stripping}, which broadly confirms the uniqueness of zero-metallicity stars in the $\psi-\Delta\Pi_1$ plane. 

Additional stars are found at $\Delta\Pi_1 \approx 60\mathrm{s}$ with elevated values of $\psi$, including KIC 9474201 ($\Delta\Pi_1 = 63.3~\mathrm{s}$, $\Delta\nu = 15.05~\mu{\rm Hz}$), KIC 9945389 ($\Delta\Pi_1 = 68.8~\mathrm{s}$, $\Delta\nu = 16.20~\mu{\rm Hz}$), KIC 7509923 ($\Delta\Pi_1 = 65.9~\mathrm{s}$, $\Delta\nu = 13.22~\mu{\rm Hz}$), and KIC 6273090 ($\Delta\Pi_1 = 73.3~\mathrm{s}$, $\Delta\nu = 16.08~\mu{\rm Hz}$), all of which may similarly indicate past mass transfer events \citep{2022A&A...659A.106D}. Given their higher masses, they can also be ruled out as Pop III candidates.

\subsection{Seismic Signatures of Interstellar Accretion Onto Pop III Stars}\label{sec:accretion}

Within the BHL framework, accretion rates onto low-mass stars moving through primordial metal-enriched ISM regions can reach $\dot{M}\sim10^{-11}-10^{-15}~M_\odot~{\rm yr}^{-1}$, potentially accumulating $M_{\rm acc.} \sim 10^{-5}-10^{-3}~M_\odot$ over a stellar lifetime. If even a modest fraction {($\leq1-10\%$ by mass)} of this accreted material consists of metals, the resulting surface metallicity could reach $Z_{\rm surf} \sim 10^{-8}-10^{-6}$, overlapping with some of the most iron-poor stars observed to date in the Galaxy (see $\S$\ref{sec:introduction}). To investigate the effects of such accretion episodes on the asteroseismic properties of low-mass Pop III stars, we computed stellar evolution models for a $0.85~M_\odot$ zero-metallicity star beginning at the ZAMS, exploring six accretion scenarios with varying accretion rates and total accreted masses as described in Table \ref{tab:accretion}. All models include full treatment of atomic diffusion and gravitational settling as described in Paper I.

\begin{deluxetable}{ccccc}[!h]
\tablecaption{Accretion parameters and resulting stellar properties. $Z_{\rm final}$ indicates the surface metal abundance at the end of the accretion phase.}
\label{tab:accretion}
\tablehead{
\colhead{$\dot{M}$} &
\colhead{$M_{\rm final}$} &
\colhead{$\Delta M$} &
\colhead{$\Delta\tau_{\rm acc}$} &
\colhead{$Z_{\rm final}$} \\
\colhead{($M_\odot\,{\rm yr}^{-1}$)} &
\colhead{($M_\odot$)} &
\colhead{($M_\odot$)} &
\colhead{} &
\colhead{}
}
\startdata
$1\times10^{-11}$ & $0.855$   & $10^{-3}$  & $500~{\rm Myr}$ & $\sim10^{-6}$ \\
$1\times10^{-13}$ & $0.855$   & $10^{-3}$  & $50~{\rm Gyr}$  & $\sim10^{-4}$ \\
$1\times10^{-13}$ & $0.85005$ & $10^{-5}$  & $500~{\rm Myr}$ & $\sim10^{-6}$ \\
$1\times10^{-14}$ & $0.85005$ & $10^{-5}$  & $5~{\rm Gyr}$   & $\sim10^{-7}$ \\
$5\times10^{-13}$ & $0.855$   & $10^{-3}$  & $10~{\rm Gyr}$  & $\sim10^{-4}$ \\
$5\times10^{-15}$ & $0.85005$ & $10^{-5}$  & $10~{\rm Gyr}$  & $\sim10^{-4}$ \\
\enddata
\end{deluxetable}

\begin{figure}
    \centering
    \includegraphics[width = \linewidth]{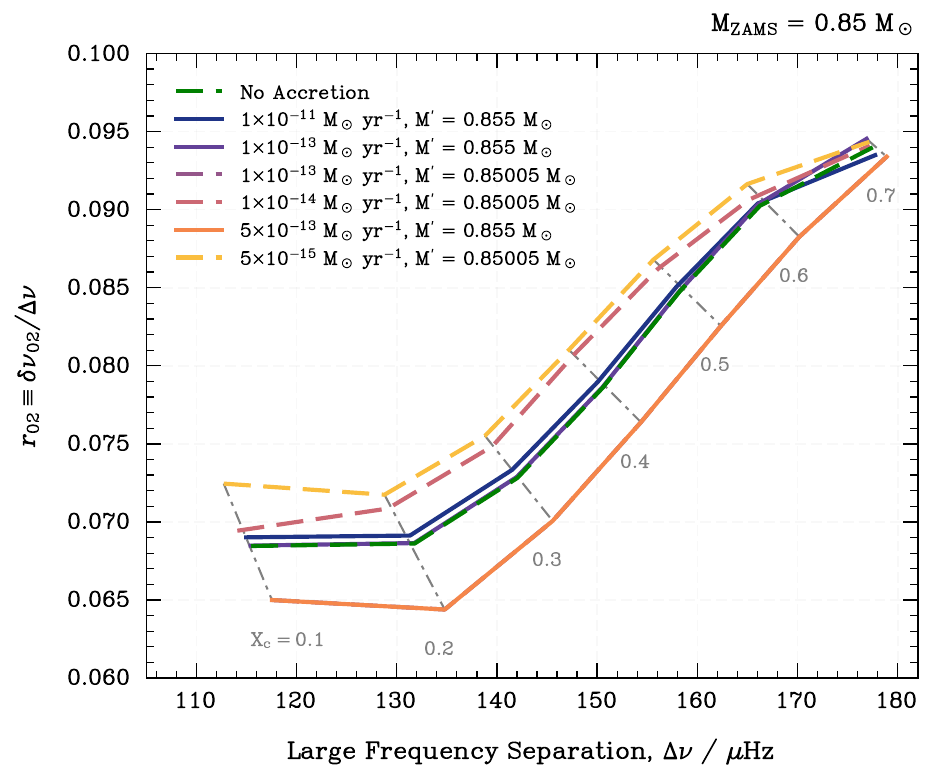}
    \includegraphics[width = \linewidth]{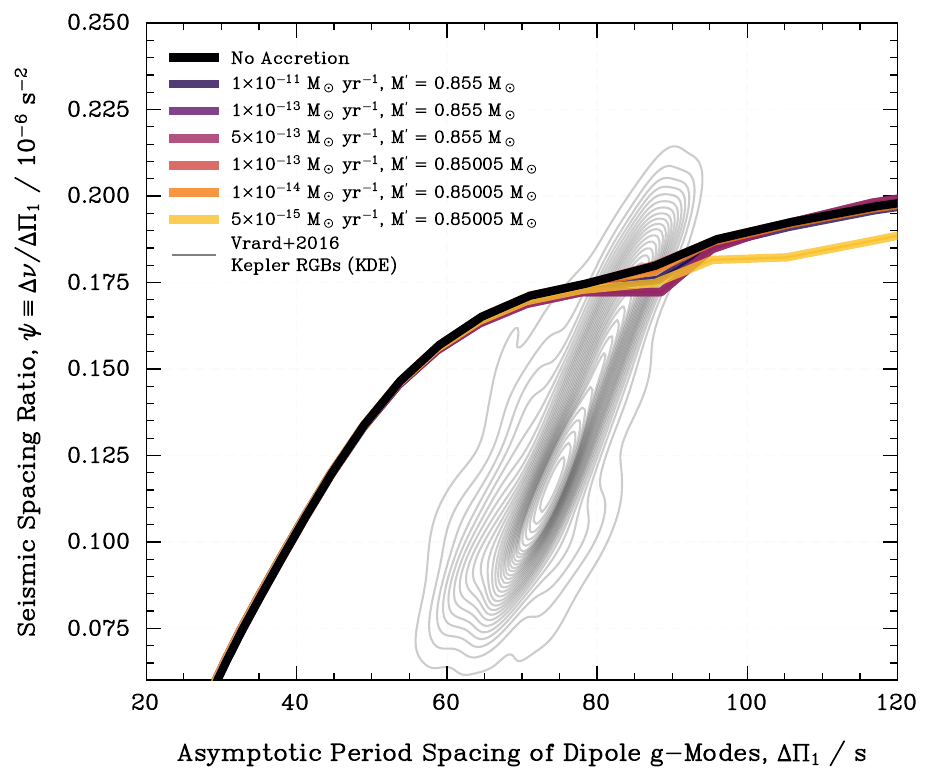}
    \caption{Asteroseismic diagnostics for ISM accretion scenarios onto a $0.85~M_\odot$ Pop III star. {\it First panel}: Small-to-large frequency separation ratio $r_{02}$  vs. $\Delta\nu$ showing subtle differences between accretion models (A--F, coloured and sorted by accretion rates) and the non-accreting reference (dashed black) during the main-sequence. Grey lines connect constant central hydrogen mass fraction $X_{\rm c}$ values ($0.1-0.7$); longer timescales enable deeper metal mixing and larger interior structural modifications. {\it Second panel}: Seismic spacing ratio $\psi$ versus $\Delta\Pi_1$ on the RGB phase. All accretion scenarios converge to nearly identical tracks (coloured with circles and sorted by the accretion rate and final mass), separated from metal-poor {\it Kepler} RGB stars (grey contours; \citealt{2016A&A...588A..87V}).}
    \label{fig:accretion}
\end{figure}

{During the main-sequence phase, we observe subtle but theoretically measurable differences in asteroseismic signatures between accretion scenarios, particularly in the small-to-large frequency separation ratio $r_{02}$ (Figure \ref{fig:accretion}, first panel), with these variations depending systematically on both the accretion rate and total accreted mass, with tracks showing distinct separation in the $r_{02}-\Delta\nu$ diagram, which reflects structural changes induced by surface pollution and diffusion. However, despite this intrinsic sensitivity, the amplitude of these acoustic oscillation signals is likely too small for detection with present-day space-based photometry missions, stemming from the low luminosity and small surface convection zones of main-sequence stars at this mass, which reduce the driving efficiency of solar-like oscillations. Ergo, individual p-modes, and therefore $r_{02}$, are unlikely to be detectable, and main-sequence asteroseismology is unlikely to provide a practical diagnostic for distinguishing polluted from pristine low-mass Pop III stars alone with present-day telescopes.}

Our critical finding emerges on the giant branch, with the accreted material undergoing efficient mixing and settling due to atomic diffusion and gravitational processes, and as the star ascends the giant branch and develops a deep convective envelope, the stellar structure returns to a state of relative homogeneity. The seismic spacing ratio $\psi$ shows no significant variation across all six accretion scenarios (Figure \ref{fig:accretion}, second panel), with all models remaining tightly clustered in the $\psi-\Delta\Pi_{1}$ plane and separated from the distribution of standard metal-poor RGB stars observed by {\it Kepler}. Such remarkable convergence demonstrates that whilst surface contamination would render these stars spectroscopically indistinguishable from extremely metal-poor (${\rm [Fe/H]}\leq-6$) stars, their internal structure and mixed-mode oscillation properties retain robust signatures of their primordial zero-metallicity cores, which are accessible with current high-precision asteroseismic observations. 

We note that these accretion scenarios represent simplified, exploratory calculations designed to bracket plausible parameter space, whereas in reality, accretion rates, total accreted masses, and accretion timescales depend sensitively on uncertain factors including the star's velocity relative to the ISM, local gas density and metallicity, and the evolutionary history of the surrounding environment. The ISM composition itself evolves as successive generations of supernovae enrich the medium, whilst the density structure of the early Universe remains poorly constrained. Furthermore, accretion may occur episodically rather than continuously, and the accreted material's composition need not match the bulk ISM metallicity. A comprehensive exploration of the full parameter space---including time-dependent accretion histories, variable metallicity profiles, and self-consistent treatment of the coupling between stellar evolution and environmental accretion---is beyond the scope of this work but represents an important and interesting avenue for future investigations.

\subsection{Seismic Signatures of Envelope Stripping in High Metallicity Stellar Models}\label{sec:stripping}

Envelope stripping via wind mass loss or Roche Lobe-overflow mass transfer in binary systems represents another potential pathway by which low-mass stars could exhibit unusual surface compositions whilst retaining signatures of their primordial origins; hence, to investigate how such stripping events affect the asteroseismic properties of initially metal-poor stars, we computed {\it artificial}\footnote{At $Z = 10^{-3}$, radiatively driven winds scale roughly as $\dot{M} \propto Z^{0.7}$ and yield $\dot{M} \leq 10^{-10}~M_\odot~{\rm yr^{-1}}$ for a $2~M_\odot$ star, i.e., far too weak to remove its envelope over the main-sequence lifetime (see \citealt{2001A&A...369..574V}). The imposed envelope-wind-stripping mechanism here, therefore, represents a controlled numerical experiment mimicking binary-induced or dynamical mass loss.} stellar evolution models for $1$, $1.5$ and $2~M_\odot$ stars with initial metallicities $Z = 10^{-3}$ in {\sc MESA}. We explored nine envelope stripping scenarios with varying residual hydrogen envelope mass: (i--iii) {\it 1.5 M$_\odot$ models} with $M_{\rm H,env} = 0.7$, $0.75$, and $0.8~M_\odot$, (iv--vi) {\it 2.0 M$_\odot$ models} with $M_{\rm H,env} = 0.5$, $0.6$, and $0.7~M_\odot$, and (vii--ix) {\it 1.0 M$_\odot$ models} with $M_{\rm H,env} = 0.7$, $0.8$, and $0.9~M_\odot$. In all cases, stripping is applied once the model reaches the main-sequence turn-off (MSTO), after which the wind-stripped star continues its evolution through the sub-giant branch and onto the red giant branch. These scenarios represent idealised mass-loss mechanisms that reduce the envelope mass whilst preserving the underlying core structure formed during main-sequence evolution with finite metallicity. 

\begin{figure}
    \centering
    \includegraphics[width = \linewidth]{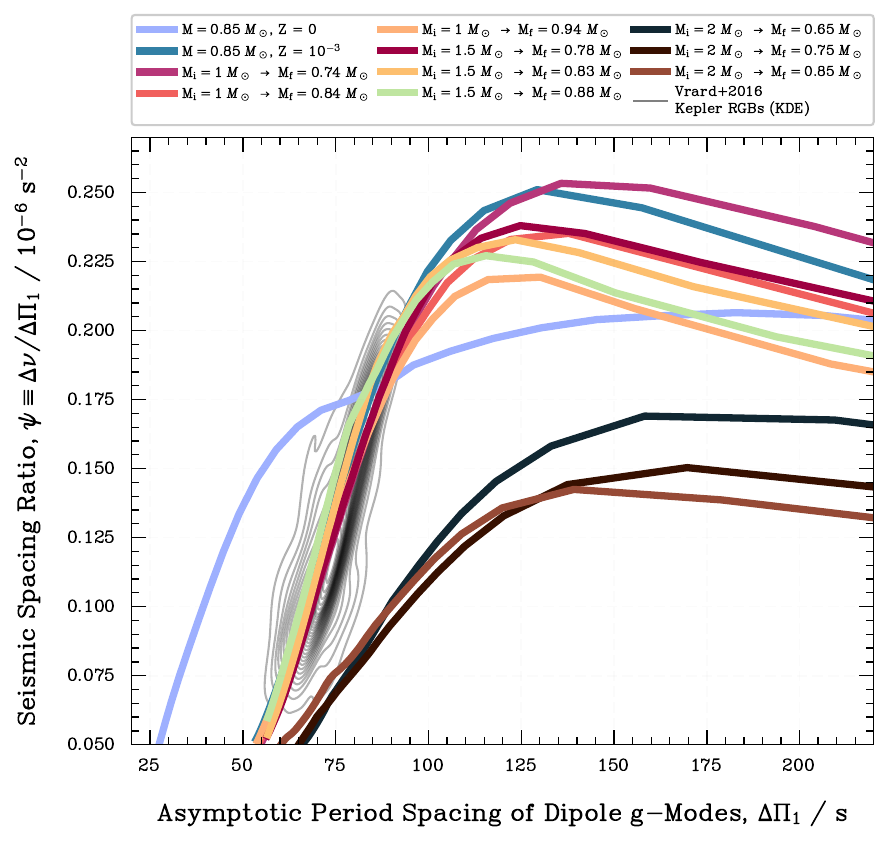}
    \caption{{Seismic spacing ratio $\psi \equiv \Delta\nu/\Delta\Pi_1$ vs. asymptotic g-mode period spacing $\Delta\Pi_1$ for envelope-stripped stellar models. For stripped models, the indicated masses correspond to the mean of the model masses along the RGB. Models starting from initial masses of $1.0~M_\odot$ (with residual H-envelope masses of $0.8~M_\odot$), $1.5~M_\odot$ (with $M_{\rm H}^\prime = 0.75~M_\odot$), and $2.0~M_\odot$ (with $M_{\rm H}^\prime = 0.7~M_\odot$) are compared to reference tracks for $0.85~M_\odot$ stars at $Z = 0$ of roughly $0.83-0.84~M_\odot$ and $Z = 10^{-3}$ (squares). Grey contours show the distribution of Kepler red-giant stars from \citet{2016A&A...588A..87V}. The stripped models exhibit systematically different seismic properties, occupying a distinct region in the parameter space.}}
    \label{fig:stripping}
\end{figure}

Figure \ref{fig:stripping} presents the $\psi-\Delta\Pi_1$ plane for envelope stripping scenarios from initial masses of $1~M_\odot$, $1.5~M_\odot$ and $2~M_\odot$ (distinguished by marker shapes: circles for $2~M_\odot$, crosses for $1.5~M_\odot$, stars for $1~M_\odot$) alongside $0.85~M_\odot$ reference tracks at $Z = 0$ and $Z = 10^{-3}$ (shown with squares). {Most envelope-stripped models are offset from the pristine low-metallicity tracks, whilst some overlap with standard metal-poor Pop II models. The most strongly stripped models, however, occupy a clearly distinct locus in the diagram, demonstrating that extreme envelope stripping can produce seismic signatures that differ from second-generation lower-mass stars, whereas more moderate stripping may yield tracks that are somehow less easily distinguishable.}

Crucially, we observe that the wind-stripped models do not converge to the $Z = 0$ track despite their reduced envelope masses and terminal-age masses of roughly $0.7-0.9~M_\odot$ approaching those of the Pop III stellar models considered in this study, with such persistent divergence demonstrating that envelope stripping of initially metal-enriched stars cannot replicate the internal structure and asteroseismic signatures of truly pristine zero-metallicity stars. Although stripping removes outer layers, core composition gradients established during main-sequence CNO-cycle (finite metallicity) versus pp-chain ($Z = 0$) burning remain intact. At $\log(L/L_\odot) \approx 1.5$, wind-stripped models exhibit He cores $22-27\%$ smaller than pristine $Z = 0$ stars ($0.22-0.24~M_\odot$ versus $0.30~M_\odot$), a consequence of the more compact convective cores produced during the shorter main-sequence lifetimes of initially more massive metal-enriched progenitors. These structural differences manifest in systematically distinct Brunt-V{\"a}is{\"a}l{\"a} frequency profiles governing g-mode propagation cavities, whereas the period spacing $\Delta\Pi_1$, i.e., an integral over the Brunt-V{\"a}is{\"a}l{\"a} frequency profile (see Equation \ref{eq:periodspacing}), differs by $27-53\%$ at similar luminosity ($51$~s for $Z = 0$ versus $66-79$~s for wind-stripped models), whilst $\Delta\nu$ (sensitive to envelope structure and mean density; see Equation \ref{eq:deltanu}) responds directly to mass loss. Consequently, the seismic spacing ratio $\psi \equiv \Delta\nu/\Delta\Pi_1$ differs by $\approx30-43\%$ ($0.140$ for $Z = 0$ versus $0.081-0.099$ for wind-stripped metal-enriched models), providing a robust asteroseismic diagnostic: envelope stripping modifies the envelope-sensitive $\Delta\nu$, but the core-sensitive $\Delta\Pi_1$ retains an indelible metallicity signature, which yields distinctive locations in the $\psi-\Delta\Pi_1$ plane. 

These results complement our findings in Section~\ref{sec:accretion}, together establishing that asteroseismology provides a robust diagnostic for distinguishing genuine Pop III stars from both accreted-polluted and wind-stripped metal-poor lookalikes. Whilst surface spectroscopy alone might struggle to differentiate these populations, the internal structure probed by mixed-mode oscillations preserves clear signatures of the star's primordial composition and evolutionary history. We also emphasise, however, some limitations of this approach that could be explored in future dedicated studies: (i) whilst our stripping is implemented through enhanced winds applied gradually over the RGB ascent (preserving realistic timescales), we do not model the detailed hydrodynamics of mass loss or account for metallicity-dependent wind prescriptions that may affect the stripping efficiency (e.g., \citealt{2005A&A...442..587V, 2021MNRAS.503..694T}); (ii) binary interactions such as tidal spin-up, orbital angular momentum transfer, and Roche lobe overflow dynamics are not included, which may also significantly alter the timing and geometry of mass loss in close binary systems \citep{2002MNRAS.329..897H, 2013A&A...554A.130L}; (iii) we consider only a single stripping episode initiated at the MSTO, whereas real systems may experience multiple phases of enhanced mass loss throughout post-main-sequence evolution; (iv) the assumption of spherical symmetry in our models neglects potential asymmetries in the mass-loss geometry that could arise in binary configurations or magnetically-channelled winds \citep{2012A&A...540A..32G, 2014ApJ...789...28P}; and (v) we do not explore the observational selection effects that would govern the detection of such wind-stripped stars in current asteroseismic surveys \citep{2014ApJ...789...28P, 2023Sci...382.1287D}.  

\subsection{Influence of Convective Efficiency on the Seismological Signatures of Zero-Metallicity Stars}\label{sec:convective}

\begin{figure}
    \centering
    \includegraphics[width = \linewidth]{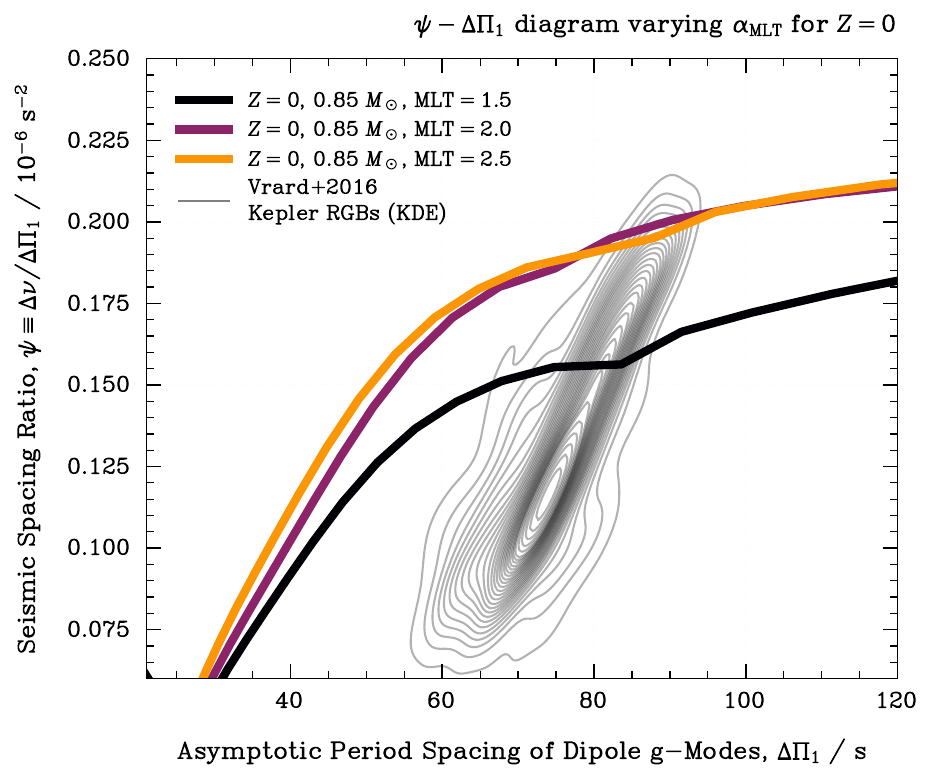}
    \caption{Seismic spacing ratio $\psi \equiv \Delta\nu/\Delta\Pi_1$ as a function of $\Delta\Pi_1$ for $0.85~M_\odot$ zero-metallicity models with varying convective mixing-length parameter $\alpha_{\rm MLT} = 1.5, 2.0, 2.5$. Grey contours show the distribution of Kepler red-giant stars from \citet{2016A&A...588A..87V}. Increasing $\alpha_{\rm MLT}$ raises $\psi$ at fixed $\Delta\Pi_1$, modestly shifting the Pop III track relative to the observed locus.}
    \label{fig:Psi_DeltaPi_MLT_Varying_Z0}
\end{figure}

Whilst metallicity proves to be the main determinant of the morphology of the $\psi-\Delta\Pi_1$ plane, variations in convective efficiency also introduce a measurable secondary modulation. In metal-free stars, their exceptionally low envelope opacity renders convection efficient even for modest values of the mixing-length parameter, and nevertheless, increasing $\alpha_{\rm MLT}$ further enhances convective heat transport, reduces the super-adiabatic gradient, and promotes slightly hotter and more compact red-giant envelopes at fixed evolutionary stage as discussed in Paper I. Such structural adjustments raise the mean density and therefore $\Delta\nu$, whilst modestly lowering $\Delta\Pi_1$ through changes to the outer g-mode cavity. The combined effect is a systematic elevation of $\psi$ with increasing $\alpha_{\rm MLT}$ along the giant branches seen in Figure \ref{fig:Psi_DeltaPi_MLT_Varying_Z0}.

{The impact becomes most apparent near the $\psi$ maximum preceding core helium ignition, where even small differences in envelope entropy and acoustic-cavity extent accumulate, shifting the $\psi$ peak to slightly lower $\Delta\Pi_1$ for larger $\alpha_{\rm MLT}$, and producing a steeper pre-ignition rise. Physically, this behaviour reflects the manner in which enhanced convective efficiency accelerates envelope cooling whilst leaving the core's delayed stratification largely intact, characteristic of the pp-chain regime in primordial stars. The consequent decoupling between envelope contraction and core entropy decline yields a subtly displaced turnover in the $\psi-\Delta\Pi_1$ locus. Although the magnitude of this effect remains subordinate to metallicity-driven differences, it nonetheless may contribute to the observed scatter among metal-poor giants and should be accounted for when attempting to identify Pop III candidates from seismic diagnostics alone.}

\section{Discussions and Conclusions}\label{sec:conclusions}

{We present an asteroseismic characterisation of low-mass, metal-free red giants, demonstrating that primordial stellar interiors imprint an unambiguous and persistent seismic fingerprint that cannot be replicated by metal-enriched evolutionary pathways. By extending the analysis presented in Paper I \citep{2026arXiv260216082F} from the main sequence through the red giant branch, we show that Pop III stars occupy a seismically distinct regime, with systematically different frequency separations, mixed-mode coupling properties, and diagnostic ratios---most notably in the $\psi-\Delta\Pi_1$ plane---arising from their fundamentally different core structures and burning histories. Crucially, these signatures remain robust even under surface pollution and envelope stripping, establishing asteroseismology as a uniquely powerful and physically grounded method for identifying and characterising relic Pop III stars as well as providing a concrete framework for targeted observational searches for the oldest stellar populations in the Galaxy.}

\begin{enumerate}
    
    \item {\it Large and Small Frequency Separations {along the Main Sequence}}. The seismic spectra of Pop III stars on the main sequence are marked by systematically depressed large frequency separations ($\Delta\nu$) and elevated separation ratios ($r_{02} \equiv \delta\nu_{02}/\Delta\nu$) relative to stars of similar mass at non-zero metallicity (Figure \ref{fig:JCD}). Such reduced $\Delta\nu$ values reflect higher mean sound speeds and lower mean densities, whilst larger $r_{02}$ values indicate sharper sound-speed gradients between the helium-rich core and hydrogen-rich envelope. Collectively, these features may provide a distinctive seismic fingerprint of metal-free stellar structure during the main-sequence of evolution. 
    
    {Moreover, the metallicity dependence of asteroseismic parameters such as $\nu_\mathrm{max}$ and $\Delta\nu$ has been explored in works including \citet{2017ApJ...843...11V, 2024ApJ...962..118Z}, which highlights potential corrections needed for accurate characterisation of metal-poor and/or Pop~III stars}.\smallskip

    \item {\it Seismic Cavity Structure and Mixed-Mode Diagnostics along the Giant Branch}. {The seismic cavity structures of Pop III stars differ systematically from metal-enriched stars, with more extended envelopes, larger helium cores, and modified Brunt–V{\"a}is{\"a}l{\"a} frequency profiles that reshape the acoustic and buoyancy cavities (see Figures \ref{fig:propagation} and \ref{fig:DeltaNuDeltaPi1}). These structural differences lead to significantly smaller asymptotic period spacings $\Delta\Pi_1$ at fixed surface properties, which become increasingly pronounced along the subgiant and red-giant branches. Such seismic signatures provide a robust diagnostic of primordial stellar interiors and may enable observational discrimination between Pop III and metal-enriched stellar populations.}

    \item {\it The $\psi-\Delta\Pi_1$ Plane}. {As shown in Figures \ref{fig:DeltaNuDeltaPi1} and \ref{fig:psi}, the parameter $\psi \equiv \Delta\nu/\Delta\Pi_1$ emerges as a sensitive probe of internal stratification and mode trapping, and exhibits a strong dependence on metallicity. This behaviour arises because $\psi$ encapsulates compensating structural dependencies: higher $\Delta\nu$ reflects increased sound speed and lower mean densities, whilst lower $\Delta\Pi_1$ traces the less stratified cores typical of metal-free evolution. The resulting turnover in $\psi$ at large $\Delta\Pi_1$ values exhibited by $Z = 0$ models reflects the slower development of mean molecular weight gradients and the weaker buoyancy frequency contrasts in Pop III interiors (see Figures \ref{fig:propagation}). Such features place primordial stars in a seismically distinct locus, separated from metal-enriched sequences. Hence, the $\psi - \Delta\Pi_1$ plane offers improved observational separation between stellar populations, being particularly powerful for red giant stars observations via long-duration space-based photometry, and could be integrated into future efforts to identify and characterise the relics of the first stellar generations.}
    
    Although sub-dominant to metallicity, mixing-length variations impart a systematic upward shift in $\psi$ and a modest turnover displacement, stressing the need to account for convective efficiency when inferring primordial candidates from the $\psi$–$\Delta\Pi_1$ plane as explored in Section~\ref{sec:convective}.

    {We further emphasise that the period spacings measured in observations, $\Delta P$, can differ from the theoretical asymptotic values, $\Delta \Pi_{1}$, computed from the Brunt–V{\"a}is{\"a}l{\"a} integral in the core, with observed $\Delta P$ being often smaller than $\Delta \Pi_{1}$ due to (i) measurement-dependent effects, as different analysis methods of mixed-mode spectra can yield systematically different spacings, and (ii) physical effects, including mode bumping, rotational splitting, and internal magnetism (e.g., \citealt{2017MNRAS.469.1360C, 2022ApJ...931..116L, 2023ApJ...946...92O, 2024MNRAS.527.6346R}), which modify g–p mode coupling. Our current models do not include rotation or magnetic fields; therefore, whilst they capture the general seismic signatures of zero-metallicity stars, reproducing observed $\Delta P$ would require both additional physical modelling and careful consideration of observational techniques.}

    {Furthermore, stripping episodes in metal-enriched models alters the envelope but cannot erase the core's metallicity imprint, i.e., wind-stripped metal-enriched stars occupy a distinct locus in the $\psi-\Delta\Pi_1$ plane, demonstrating that asteroseismology can unambiguously distinguish genuine Pop III stars from stripped metal-poor impostors.} \smallskip

    \item {\it Observational Outlook and Prospects for Seismic Detection}. We investigated surface contamination episodes in low-mass Pop III models and found that moderate levels of external or self-pollution have a negligible impact on their internal structure and pulsation properties of zero-metallicity stars (see Section~\ref{sec:accretion}). Consequently, whilst such stars are unlikely to be spectroscopically distinguishable once polluted, their internal structure and oscillation modes remain essentially unaltered by surface enrichment. Combined asteroseismic diagnostics, specifically $\psi$, therefore offer another robust means of identifying relic Pop III stars (see Figure \ref{fig:DeltaNuDeltaPi1}). \smallskip

    {Pop III stellar evolutionary timescales are strongly dependent on both mass and metallicity, which directly impacts their present-day detectability. As demonstrated in Paper I, low-mass Pop III stars evolve more slowly than their metal-enriched counterparts since hydrogen burning proceeds predominantly via the pp chain and convective mixing is limited at zero metallicity, yielding main-sequence lifetimes that can exceed the Hubble time for $M \lesssim 0.8~M_\odot$. This behaviour can be characterised by a simple power-law scaling, $\Gamma_{\rm MS,~Pop III} \propto \left(M/M_\odot\right)^{-3.2}$, illustrating the steep decrease in lifetime with increasing mass. Conversely, massive Pop III stars evolve rapidly and end their lives as supernovae on million-year timescales, preventing their survival to the present epoch. Cosmological simulations and semi-analytic models of early star formation further suggest that low-mass Pop III stars can survive to the present day, predominantly in the Galactic halo, though their number densities are extremely low (e.g., \citealt{2015MNRAS.447.3892H, 2016MNRAS.462.1307S, 2016ApJ...820...59K, 2020ApJ...901...16D}. These studies further indicate that fragmentation in primordial mini-halos can produce multiple low-mass stars per halo, even if the overall Pop III initial mass function is top-heavy, providing a physical basis for the existence of long-lived survivors. Consequently, the evolutionary phases discussed in this work pertain exclusively to these rare, long-lived Pop III stars, and any observational detection would probe a physically plausible but extremely sparse population, consistent with current constraints from metal-poor halo stars.}

    All of these findings support the development of dedicated seismic surveys of metal-poor stellar populations and incorporate asteroseismic criteria into searches for primordial stars using upcoming facilities (and concepts) such as the PLAnetary Transits and Oscillations of stars (PLATO/ESA; \citealt{2025ExA....59...26R}) for asteroseismology, the ArmazoNes high Dispersion Echelle Spectrograph at the Extremely Large Telescope (ANDES-ELT/ESO; \citealt{2021Msngr.182...27M}) for high-resolution spectroscopy follow-up, the wide-field space telescope Xuntian/CSST \citep{2025arXiv250115023G} for deep imaging and spectroscopy, and next-generation ultraviolet observatories such as Large Ultraviolet Optical Infrared Surveyor (LUVOIR/NASA; \citealt{2019arXiv191206219T}) and Cosmic Evolution Through UV Spectroscopy (CETUS/NASA; \citealt{2017AAS...22923827H}) high-sensitivity UV spectroscopy.

\end{enumerate}

\begin{acknowledgments}  
    
    We warmly thank Sarbani Basu, Pratik Gandhi, and Saskia Hekker for insightful discussions that greatly enriched this work, and we are grateful to the anonymous referee for constructive comments that significantly improved the clarity and depth of the manuscript. We thank the Yale Center for Research Computing (YCRC) for providing access to the Grace computing cluster used in this study. TF acknowledges support from the Yale Graduate School of Arts and Sciences. EF thanks the Yale Center for Astronomy and Astrophysics Prize Fellowship. CJL acknowledges support from a Gruber Science Fellowship and from NSF grant AST-2205026.

\end{acknowledgments}

%\begin{contribution} \end{contribution}

\software{MESA \citep{2011ApJS..192....3P, 2013ApJS..208....4P, 2015ApJS..220...15P, 2018ApJS..234...34P, 2019ApJS..243...10P, 2023ApJS..265...15J}, GYRE \citep{2013MNRAS.435.3406T, 2018MNRAS.475..879T}.}

\bibliography{bib}{}
\bibliographystyle{aasjournalv7}

\end{document}